\pdfoutput=1

\documentclass{emulateapj}

\slugcomment{Accepted to The Astrophysical Journal}

\shorttitle{PERIODIC OPTICAL VARIABILITY OF RADIO DETECTED ULTRACOOL DWARFS}
\shortauthors{HARDING ET AL.}

\begin{document}


\title{PERIODIC OPTICAL VARIABILITY OF RADIO DETECTED ULTRACOOL DWARFS}


\author{Leon K. Harding\altaffilmark{1, 2, 6},  Gregg Hallinan\altaffilmark{2}, Richard P. Boyle\altaffilmark{3}, Aaron Golden\altaffilmark{1,5}, Navtej Singh\altaffilmark{1}, Brendan Sheehan\altaffilmark{1}, Robert T. Zavala\altaffilmark{4} and Ray F. Butler\altaffilmark{1}}

\affil{\altaffilmark{1}Centre for Astronomy, National University of Ireland, Galway, University Road, Galway, Ireland}

\affil{\altaffilmark{2}Cahill Center for Astrophysics, California Institute of Technology, 1200 E. California Blvd., MC 249-17, Pasadena, CA 91125, USA}

\affil{\altaffilmark{3}Vatican Observatory Research Group, Steward Observatory, University of Arizona, Tucson, AZ 85721, USA}

\affil{\altaffilmark{4}United States Naval Observatory, Flagstaff Station, Flagstaff, AZ, USA}

\affil{\altaffilmark{5}Department of Genetics (Computational Genetics), Albert Einstein College of Medicine, Bronx NY 10461, USA}

\altaffiltext{6}{Now at Caltech: lkh@astro.caltech.edu}








\begin{abstract}
A fraction of very low mass stars and brown dwarfs are known to be radio active, in some cases producing periodic pulses. Extensive studies of two such objects have also revealed optical periodic variability and the nature of this variability remains unclear. Here we report on multi-epoch optical photometric monitoring of six radio detected dwarfs, spanning the $\sim$M8 - L3.5 spectral range, conducted to investigate the ubiquity of periodic optical variability in radio detected ultracool dwarfs. This survey is the most sensitive ground-based study carried out to date in search of periodic optical variability from late-type dwarfs, where we obtained 250 hours of monitoring, delivering photometric precision as low as $\sim$0.15\%. Five of the six targets exhibit clear periodicity, in all cases likely associated with the rotation period of the dwarf, with a marginal detection found for the sixth. Our data points to a likely association between radio and optical periodic variability in late-M/early-L dwarfs, although the underlying physical cause of this correlation remains unclear. In one case, we have multiple epochs of monitoring of the archetype of pulsing radio dwarfs, the M9 TVLM 513-46546, spanning a period of 5 years, which is sufficiently stable in phase to allow us to establish a period of 1.95958 $\pm$ 0.00005 hours. This phase stability may be associated with a large-scale stable magnetic field, further strengthening the correlation between radio activity and periodic optical variability. Finally, we find a tentative spin-orbit alignment of one component of the very low mass binary LP 349-25.
\end{abstract}





\section{INTRODUCTION}\label{intro}


\par Beyond spectral type $\gtrsim$M7 (ultracool dwarfs), H$\alpha$ and X-ray luminosities drop sharply, signaling that chromospheric and coronal heating becomes less efficient, even in the presence of rapid rotation \citep{mohanty03,west04,reiners08,west09}. Despite this reduction in quiescent emission, a number of H$\alpha$ and X-ray flares have been detected, indicating that chromospheric and coronal activity is indeed present \citep{reid99,gizis00,rutledge00,liebert03,fuhrmeister04,rockenfeller06}. Surprisingly, given the absence of quiescent emission at higher energies, \citet{berger01} reported persistent radio emission from LP 944-20 (M9) - the first detection of radio emission from a brown dwarf, orders of magnitude higher than the expected flux \citep{gudel93}.

\par To date, quiescent radio emission has been detected from ten ultracool dwarfs \citep{berger01,berger02,berger05,burgasser05,osten06,berger06,phanbao07,hallinan06,hallinan07,antonova07,berger09,route12}. Probably the most surprising aspect of this radio activity, has been the detection of periodic 100\% circularly polarized pulses  \citep{hallinan07,hallinan08,berger09}. Observations by \citet{hallinan07} of TVLM 513-46546 (henceforth TVLM 513), reveal electron cyclotron maser (ECM) emission as the mechanism responsible for these 100\% circularly polarized periodic pulses, implying kilogauss (kG) magnetic field strengths in a large-scale stable magnetic field configuration. This is consistent with the confirmation of kG magnetic field strengths for ultracool dwarfs via Zeeman broadening observations \citep{reiners07}. Although these observations confirmed the ECM process to be the cause of the polarized periodic emission, it is still unclear as to which mechanism (incoherent or coherent) is driving the quiescent component of the radio emission, and incoherent gyrosynchrotron emission has alternatively been invoked \citep{berger06,osten06}.  

\begin{deluxetable*}{lccccccccc}
\tablecaption{Summary of Campaign Sample Properties}

\tablehead{
\colhead{Source} & \colhead{SpT} & \colhead{Distance} & \colhead{I (mag)} & \colhead{log} & \colhead{\textit{v} sin \textit{i}} & \colhead{Lithium?} & \colhead{Est. Mass} & \colhead{References} & \colhead{Radio}
 \\
& & & &($L_{bol}/L_{\odot}$) & & & $M_{\rm tot}$ & & Disc. \\
& & (pc) & & & (km s$^{-1}$) & & ($M_{\odot}$) & & Ref.\\ \\
 \colhead{(1)} & \colhead{(2)} & \colhead{(3)} & \colhead{(4)} & \colhead{(5)} & \colhead{(6)} & \colhead{(7)} & \colhead{(8)} & \colhead{(9)} & \colhead{(10)}
}
\startdata
LP 349-25AB 	    & $\sim$M8+M9$^{\dag}$ & 13.10 $\pm$ 0.28 & 12.40 & -3.19; -3.34 & 55 $\pm$ 2; 83 $\pm$ 3 & No & 0.121 $\pm$ 0.009 & 1-3 & 21\\

2M J0746AB			& L0+L1.5 & 12.20 $\pm$ 0.05 & 15.03 & -3.64; -3.77 & 19 $\pm$ 2; 33 $\pm$ 2 & No & 0.151 $\pm$ 0.003 & 2, 4-6 & 22\\

LSR J1835 			& M8.5 & $\sim$6.0 & 12.90 & -3.51 & 50 $\pm$ 5 & ? & $<$0.083? & 7-9 & 23\\

TVLM 513 			& M9 & $\sim$10.5 & 15.10 & -3.65 & $\sim$60 & No & $>$0.06 & 4, 10-13 & 24\\

BRI 0021			& M9.5 & $\sim$11.5 & 15.02 & -3.40 & $\sim$34 & No & $<$0.06 & 11, 14-17 & 24\\

2M J0036			& L3.5 & $\sim$8.8 & 16.05 & -3.98 & $\sim$37 & No & 0.06 - 0.074 & 4, 5, 18-20 & 25

\enddata

\tablecomments{Column (1) Campaign source. (2) Spectral type. (3) Distance (in parsecs). (4) Magnitude as measured in the Johnson I-band. (5) Bolometric luminosity. (6) Rotation velocity (in kilometers second$^{-1}$). (7) Lithium present in previous spectroscopic studies? (8) Estimated total system mass (in solar mass). (9) Observational study references (corresponding to list below). (10) Radio emission discovery references. The campaign targets are abbreviated as follows: 2MASSW J0746425+200032 (henceforth 2M J0746); LSR J1835+3259 (henceforth LSR J1835); TVLM 513-46546 (TVLM 513); BRI 0021-0214 (henceforth BRI 0021) and 2MASS J00361617+1821104 (henceforth 2M J0036). $^{\dag}$LP 349-25 may be either M7.5+M8.5 or M8+M9 as outlined by \citet{forveille05}.\\ \\ References. --- (1) \citet{gatewood09}. (2) \citet{forveille05} \& \citet{konopacky10,konopacky12}. (3) \citet{basri05,reiners09}. (4) \citet{dahn02}. (5) \citet{vrba04}. (6) \citet{bouy04}. (7) \citet{reid03}. (8) \citet{berger08}. (9) \citet{hallinan08}. 10) \citet{tinney93,tinney95}. (11) \citet{leggett01}. (12) \citet{basri01}. (13) \citet{reid02}. (14) \citet{reid99}. (15) \citet{mohanty03}. (16) \citet{reiners09}. (17) \citet{chabrier00}. (18) Average of \citet{jones05} \& \citet{osorio06}. (19) \citet{reid00}. (20) \citet{hallinan08} based on work of \citet{reid00,vrba04}. (21) \citet{phanbao07}. (22) \citet{antonova08}. (23) \citet{berger06}. (24) \citet{berger02}. (25) \citet{berger05}.}\label{table_target_properties}

\end{deluxetable*}

\par Ultracool dwarfs have also exhibited periodic variability in the optical regime. These investigations have yielded both optical and infrared (IR) variability, where modulation at the expected rotation period has been found in various studies \citep{clarke02a,koen06,rockenfeller06,lane07,littlefair08}. Aperiodic variability, as well as periodic modulations on time-scales not associated with rotation, have been inferred \citep{gelino02,lane07,maiti07}. Typically, this variability has been attributed to magnetic spots on the surface of the dwarf, or the presence of atmospheric dust, or indeed both. For higher temperature ultracool dwarfs (specifically late-M and early-L dwarfs), the presence of magnetic spots and other magnetic related activity, as seen for earlier M-dwarfs, may be present \citep{rockenfeller06,lane07}. \citet{littlefair08} reported sinusoidal variability of the M9 dwarf TVLM 513, with a period of $\sim$2 hours - a period consistent with the radio pulsing and optical periodicity previously obtained by \citet{hallinan06} and \citet{lane07}, respectively. However, their light curves (Sloan $g^{\prime}$ and Sloan $i^{\prime}$) were anticorrelated, which seemed to refute the proposed model of starspots at that time as the cause for the optical variability. Instead they argued that this anticorrelated signal was likely due to photospheric dust coupled with stellar rotation. Indeed, magnetic activity, as signaled by H$\alpha$, decreases further after the M/L transition \citep{west04}, therefore in most cases optical variability has been attributed to the expected presence of dust in the dwarf's atmosphere \citep{jonesmundt01,martin01,gelino02,enoch03,maiti07,littlefair08,goldman08,clarke08}.

\par It is notable that two of the ultracool dwarfs found to be periodically variable in the optical \citep{lane07}, are also known to be members of the small sample known to be pulsing radio sources \citep{hallinan07,hallinan08}. Motivated by this, we have commenced a campaign to investigate whether optical periodic variability is a signature property of radio detected ultracool dwarfs. To this end, we employed the custom developed GUFI mk.II photometer, as well as the VATT 4K CCD Imager on the 1.83 m Vatican Advanced Technology Telescope (VATT)\footnote[7]{The Vatican Advanced Technology Telescope (VATT) telescope facility is operated by the Vatican Observatory, and is part of the Mount Graham International Observatory.}, to photometrically monitor all of the radio emitting ultracool dwarfs observable from the VATT site. Throughout the campaign, data were also obtained from the 1.0 m and 1.55 m telescopes at the USNO\footnote[8]{Information regarding the United States Naval Observatory (USNO) telescopes can be found here: http://www.usno.navy.mil/USNO.}, as well as the 1.52 m telescope at the Loiano Observatory in Bologna, Italy.

\section{SAMPLE}\label{objects}

\par In the following sections we discuss each target with respect to any previous radio and optical emission. A list of the campaign sample, as well as a summary of individual target information, is shown in Table~\ref{table_target_properties}. These are categorized in order of ascending spectral type; we also outline details of the respective observation campaigns. Our target sample consists of those dwarfs which have been previously detected as radio sources and are visible from the VATT observatory site. The general capabilities of each detector used for the campaign are outlined in \S~\ref{optical_obs}. Dwarfs which have been detected as optically variable sources in other work were also included for verification, and to assess the stability of these optical signals over time scales of years.

\subsection{Binary Systems}

\par We selected two very low mass binary stars at the M/L transition for our campaign - LP 349-25 and 2M J0746. These objects were of particular interest, since they are the only binary dwarfs reported thus far to exhibit radio emission in the very low mass (VLM) binary regime \citep{phanbao07,antonova08,osten09,berger09}, defined to be $M_{\rm tot}$ $\leq$0.185 $M_{\odot}$ \citep{close03}. Furthermore, both objects were subject to high-precision dynamical mass measurements \citep{dupuy10,konopacky10}, and more recently a large campaign was carried out to establish the \textit{individual} rotational velocities of each binary component \citep{konopacky12} - the first resolved AO measurements of this kind. Based on these dynamical mass and rotational velocity measurements, an accurate period of rotation provides the means of assessing the system's orbital coplanarity. Moreover, a range of radii can also be estimated. Importantly, \citet{harding13} have recently reported alignment of the spin-orbital axes of 2M J0746AB. Thus, the discovery of a rotation period from the binary LP 349-25 has allowed us to investigate this possible alignment for another VLM system.

\subsubsection{LP 349-25AB (M8V+M9V)}\label{lp349}

\par The M tight binary dwarf LP 349-25 was reported as a quiescent radio source by \citet{phanbao07} and \citet{osten09}; however no radio pulsing has been found thus far. More recently, a rotational velocity study carried out by \citet{konopacky12} of individual components of very low mass binaries, including LP 349-25, yielded a \textit{v} sin \textit{i} of 55 $\pm$ 2 km s$^{-1}$ and 83 $\pm$ 3 km s$^{-1}$ for LP 349-25A and LP 349-25B, respectively. Under the assumption of a rotational axis which is orthogonal to the orbital plane \citep{hale94}, the inferred equatorial velocities are $\sim$62 km s$^{-1}$ and $\sim$95 km s$^{-1}$, respectively. This would make LP 349-25B the fastest rotating low mass object yet discovered.

\par Thus far, no optical variability has been detected for LP 349-25. We therefore chose to monitor the binary to investigate the presence of optical variability. We used VATT R-band and I-band broadband filters for observations with GUFI mk.II over the course of three separate epochs, for a total of $\sim$64 hours, spanning $\sim$1.2 years.

\subsubsection{2MASSW J0746425+200032AB (L0+L1.5) }\label{j0746}

\par 2M J0746 is an L dwarf binary with a separation of $\sim$2.7 AU \citep{reid01}. The first detection of confirmed radio emission was reported by \citet{antonova08} during a 2 hour observation, and following this observation, \citet{berger09} reported periodic radio emission of 2.07 $\pm$ 0.002 hours, as well as quasi-sinusoidal periodic H$\alpha$ emission with the same period. \citet{berger09} proposed that the source of the periodicity in both cases was coming from the same component of the binary, that of 2M J0746A, and calculated a magnetic field strength of $\sim$1.7 kG, which was in agreement with \citet{antonova08}.

\par Recently, \citet{konopacky12} reported the first resolved \textit{v} sin \textit{i} measurements of the system. They measure a \textit{v} sin \textit{i} of 19 $\pm$ 2 km s$^{-1}$, and 33 $\pm$ 2 km s$^{-1}$, for 2M J0746A and 2M J0746B, respectively. Previously, in terms of rotation period measurement in optical photometry, there were only rough estimates based on unresolved \textit{v} sin \textit{i} data (with reported periods of 1.84 - 5.28 hours, see \citet{jones04}), as well as some photometric variability which was detected by \citet{clarke02b}, showing weak evidence of periodicity of a few hours. \citet{harding13} reported a period of 3.32 $\pm$ 0.15 hours for 2M J0746A, inferring that \citet{berger09} in fact detected the secondary in the radio. This refuted the claimed radius of 0.78 $\pm$ 0.1 $R_{J}$ for 2M J0746A, which \citet{harding13} demonstrate to be 0.99 $\pm$ 0.03 $R_{J}$.

\par A total of $\sim$62 hours of multiple epoch I-band observations were obtained over $\sim$2 years to investigate the long-term behavior of the optical variability on time scales of years. These observations were taken with the VATT 4K system as well as the GUFI mk.II photometer.

\vspace{1.2cm}

\subsection{Single Systems}

\subsubsection{LSR J1835+3259 (M8.5)}\label{lsr}

\par The ultracool dwarf LSR J1835 is a rapid rotator with a \textit{v} sin \textit{i} of 50 $\pm$ 5 \citep{berger08}. \citet{berger06} detected radio emission from LSR J1835 during a $\sim$2 hour observation, and proposed incoherent gyrosynchrotron radiation was responsible with an associated field strength of $<$30 G. \citet{hallinan08} later observed the dwarf for 11 hours using the VLA, and reported persistent 100\% circularly polarized coherent pulses of radio emission with a period of 2.84 $\pm$ 0.01 hours, which they attributed to the dwarf's rotation period. They argue in favor of electron cyclotron maser (ECM) emission as the dominant source of the pulsed radio emission, requiring magnetic fields of $\sim$3 kG. 

\par Based on the above radio activity of LSR J1835, we decided to further investigate the presence of such variability at optical wavelengths, and whether it was periodic in nature like the optical periodic variability presented by \citet{lane07} for the M9 dwarf TVLM 513. We conducted observations over a period of $\sim$3 years, encompassing three separate epochs. Initial epochs were taken as test data only for the GUFI mk.I system in July 2006 in the Johnson I-band, using the 1.52 m telescope in Loiano, Bologna, Italy. We also include Johnson I-band and R-band data from the USNO 1.55 m telescope in Flagstaff, Arizona, obtained by group members in September 2006. Finally, we observed the dwarf in the VATT I-band with the GUFI mk.II system on the 1.83 m VATT telescope, Mt. Graham, Arizona, to confirm its periodic nature in June 2009. The three epochs contain $\sim$33 hours of observations on source.

\subsubsection{TVLM 513-46546 (M9)}\label{tvlm}

\par TVLM 513 is one of the most rapidly rotating ultracool dwarfs discovered thus far with a rotation rate of $\sim$60 km s$^{-1}$ \citep{basri01}. All the same, only weak levels of H$\alpha$ have been found in its spectrum \citep{martin94, reid01, mohanty03}, with no X-ray detections reported so far. 

\par \citet{berger02} and \citet{osten06} detected transient radio emission from TVLM 513, however no obvious flaring was found. \citet{hallinan06} then reported persistent periodic radio emission with a period of $\sim$2 hours. Following this, \citet{hallinan07} revealed periodic bursts of radio emission with a period of $\sim$1.96 hours - confirming the presence of kG magnetic field strengths based on broadband, ECM coherent radio emission. These observations were conducted simultaneously to a photometric monitoring campaign by \citet{lane07}, who also detected a periodic signal of $\sim$1.96 hours in photometric I-band data (attributed to magnetic spots), establishing that the periodicity was due to the rotational modulation of the star, as put forward by \citet{hallinan06}. However, \citet{littlefair08} instead propose that atmospheric dust was responsible, after reporting anti-correlated Sloan \textit{g$^{\prime}$} and \textit{$i^{\prime}$} periodic variability of the M9 dwarf. Periodic H$\alpha$ and H$\beta$ variability has also be reported \citep{berger08}, perhaps indicating the presence of localized heating in the dwarf's chromospheric regions.

\par We observed TVLM 513 in optical photometric VATT I-band observations with GUFI mk.II on VATT in June 2009, in addition to three additional I-band epochs in February and April 2011, and in May 2011 using the the VATT 4K CCD and a Sloan \textit{$i^{\prime}$} filter. Data taken by members of the group using the USNO 1.0 m telescope is also included, from an epoch in 2008, and earlier VATT data obtained in 2006. This baseline therefore extends for $\sim$5 years encompassing $\sim$53 hours of data.

\subsubsection{BRI 0021-0214 (M9.5)}\label{bri}

\par In a campaign investigating magnetic activity in ultracool dwarfs, \citet{berger10} found steady and variable H$\alpha$ emission from BRI 0021 on a $\sim$0.5 - 2 hour time-scale, albeit no detected radio emission, despite previous low-level detections of radio emission \citep{berger02}. \citet{reid99} also reported a weak H$\alpha$ flare. Other optical variability has been reported by \citet{martin01}, who find I-band variability during multi-epoch photometric observations with some evidence of periodicity ($\sim$20 hours and $\sim$4.8 hours) in their analysis. They argue that since the dwarf appeared to have low levels of magnetic activity, the variability was probably not due to surface spots, but rather due to dust clouds in the dwarf's atmosphere -  since the presence of silicate and iron clouds are expected based on the dwarf's spectrum \citep{chabrier00}. It is a rapidly rotating dwarf with a \textit{v} sin \textit{i} $\approx$ 34 km s$^{-1}$ \citep{reid99,mohanty03}.

\par Based on the above radio and optical studies, we observed the dwarf in broadband optical photometry with GUFI mk.II, and obtained $\sim$28 hours of I-band data over three epochs of $\sim$1.2 years of separation.

\subsubsection{2MASS J00361617+1821104 (L3.5)}\label{j0036}

\par 2M J0036 is a radio active ultracool dwarf with rotation velocity estimates of $\sim$15 km s$^{-1}$, 38 km s$^{-1}$ and 36 km s$^{-1}$, respectively, based on a number of studies \citep{schweitzer01,jones05,osorio06}. 

\par \citet{berger05} confirmed the presence of highly variable, periodic radio emission, with a period of $\sim$3 hours. This level of radio emission violated the G\"udel-Benz relationship by many orders of magnitude (see \citet{gudel93}). They interpret the emission as incoherent gyrosynchrotron radiation, with a corresponding magnetic field strength of 175 G. However, \citet{hallinan08} reported 2M J0036 to be once again a persistent source of radio emission, and based on the periodic presence of 100\% circularly polarized emission, ruled out gyrosynchrotron radiation and confirmed ECM emission as the mechanism responsible for the pulsed radio emission. This required a magnetic field strength of at least 1.7 kG, which was the first confirmation of kG magnetic field strengths for an L dwarf.

\par Prior to these observations, \citet{lane07} conducted photometric I-band observations of 2M J0036, and found the dwarf to be photometrically variable, with a periodicity of $\sim$3 hours, arguing that magnetic spots on the surface of the dwarf, coupled with the rotation of the star, were a likely source of the periodicity. Some evidence of aperiodic variability was also present, which they attribute to dust clouds in the cooler L dwarf atmosphere.

\par We chose to observe 2M J0036 in optical photometry in the same optical band as \citet{lane07} to determine whether the optical periodicity was present over time-scales of years. We used GUFI mk.II on VATT at I-band wavelengths, for two nights in December 2010, for a total of $\sim$10 hours.

\begin{deluxetable*}{lcccccccccc}
\tablecaption{Observation Details}

\tablehead{
\colhead{Source} & \colhead{Epochs} & \colhead{Total Time} & \colhead{Date} & \colhead{Length} & \colhead{Exp.} & \colhead{Band} & \colhead{Readout} & \colhead{Amp} & \colhead{Refs} & \colhead{Telescope} \\
& & /Baseline & of Obs. & of Obs. & Time & & Rate & & & / Inst.\\
& (\#) & ($\sim$hrs; yrs) & (UT) & ($\sim$hrs) & (s $\times$ coadd) & & (MHz) && (\#) & \\ \\
 \colhead{(1)} & \colhead{(2)} & \colhead{(3)} & \colhead{(4)} & \colhead{(5)} & \colhead{(6)} & \colhead{(7)} & \colhead{(8)} & \colhead{(9)} & \colhead{(10)} & \colhead{(11)}
}
\startdata
LP 349-25AB 			& 3 & 64; 1.2 & 2009 Sept 22 & 7.2 & 5 $\times$ 24 & I & 1 & Conv. & 5 & VATT/GUFI \\

				& & & 2009 Sept 26 & 4.0 & 5 $\times$ 24& I & 1 & Conv.& 5 & VATT/GUFI\\
				
				& & & 2010 Oct 9 & 4.0 & 5 $\times$ 12& I & 1 & Conv.& 5 & VATT/GUFI\\

				& & & 2010 Oct 10 & 6.4 & 5 $\times$ 12& I & 1 & Conv. & 5 & VATT/GUFI\\

				& & & 2010 Oct 11 & 5.2 & 5 $\times$ 12& I & 1 & Conv. & 5 & VATT/GUFI\\

				& & & 2010 Oct 12 & 5.5 & 5 $\times$ 12& I & 1 & Conv.& 5 & VATT/GUFI\\

				& & & 2010 Oct 13 & 6.5 & 5 $\times$ 12& I & 1 & Conv. & 5 & VATT/GUFI\\

				& & & 2010 Oct 14 & 7.0 & 5 $\times$ 12& I & 1 & Conv. & 5 & VATT/GUFI\\

				& & & 2010 Oct 15 & 6.0 & 5$\times$ 12 & R & 1 & Conv.& 4 & VATT/GUFI\\

				& & & 2010 Nov 16 & 7.3 & 5 $\times$ 12& I & 1 & Conv. & 5 & VATT/GUFI\\

				& & & 2010 Nov 27 & 5.0 & 5 $\times$ 12& I & 1 & Conv. & 5 & VATT/GUFI\\ 
\hline\\
2M J0746AB			& 4 & 62; 2 & 2009 Jan 25 & 6.0 & 25 $\times$ 1& I & ... & Conv. & 20 & VATT/4K \\

				& & & 2009 Jan 26 & 6.8 & 25 $\times$ 1& I & ... & Conv.& 15 & VATT/4K\\
				
				& & & 2009 Jan 28 & 7.4 & 25 $\times$ 1& I & ... & Conv.& 19 & VATT/4K\\

				& & & 2010 Feb 19 & 4.5 & 5 $\times$ 12 & I & 1 & Conv. & 5 & VATT/GUFI\\

				& & & 2010 Feb 20 & 4.0 & 5 $\times$ 12& I & 1 & Conv. & 6 & VATT/GUFI\\

				& & & 2010 Nov 13 & 4.6 & 5 $\times$ 12& I & 1 & Conv.& 6 & VATT/GUFI\\

				& & & 2010 Nov 14 & 5.5 & 5 $\times$ 12& I & 1 & Conv. & 5 & VATT/GUFI\\

				& & & 2010 Dec 2 & 6.0 & 5 $\times$ 12& I & 1 & Conv. & 6 & VATT/GUFI\\

				& & & 2010 Dec 12 & 3.0 & 5 $\times$ 12& I & 1 & Conv.& 6 & VATT/GUFI\\

				& & & 2010 Dec 13 & 6.8 & 5 $\times$ 12& I & 1 & Conv. & 6 & VATT/GUFI\\

				& & & 2010 Dec 14 & 7.0 & 5 $\times$ 12& I & 1 & Conv. & 6 & VATT/GUFI\\ 
\hline \\
LSR J1835		& 3 & 33; 3 & 2006 Jul 17 & 7.0 & 5 $\times$ 12 & I & 1 & Conv. & 5 & Loiano/GUFI \\

				& & & 2006 Jul 20 & 6.5 & 5 $\times$ 12& I & 1 & Conv.& 5 & Loiano/GUFI\\
				
				& & & 2006 Sept 22 & 3.6 & 30 $\times$ 2& I & ... & ... & 10 & USNO/Tek2k\\

				& & & 2006 Sept 24 & 3.0 & 30 $\times$ 2& R & ... & ... & 10 & USNO/Tek2k\\

				& & & 2009 Jun 11 & 2.2 & 5 $\times$ 12 & I & 1 & Conv. & 5 & VATT/GUFI\\

				& & & 2009 Jun 13 & 4.0 & 5 $\times$ 12& I & 1 & Conv. & 5 & VATT/GUFI\\

				& & & 2009 Jun 16 & 4.0 & 5 $\times$ 12& I & 1 & Conv.& 4 & VATT/GUFI\\

				& & & 2009 Jun 30 & 3.0 & 5 $\times$ 12& I & 1 & Conv. & 5 & VATT/GUFI\\ 
\hline \\
TVLM 513 		& 6 & 53; 5 & 2006 May 21 & 4.8 & 30 $\times$ 3 & I & ... & ... & 6 & VATT/2K \\

				& & & 2008 Jun 17 & 6.0 & 60 $\times$ 2.5& I & ... & ...& 10 & USNO/new2k\\
				
				& & & 2009 Jun 12 & 3.6 & 5 $\times$ 12& I & 1 & Conv.& 5 & VATT/GUFI\\

				& & & 2009 Jun 13 & 4.1 & 5 $\times$ 12 & I & 1 & Conv. & 5 & VATT/GUFI\\

				& & & 2009 Jun 16 & 4.0 & 5 $\times$ 12& I & 1 & Conv. & 6 & VATT/GUFI\\

				& & & 2011 Feb 18 & 3.5 & 5 $\times$ 12& I & 1 & Conv.& 6 & VATT/GUFI\\

				& & & 2011 Feb 25 & 4.3 & 5 $\times$ 12& I & 1 & Conv. & 5 & VATT/GUFI\\

				& & & 2011 Apr 12 & 7.0 & 5 $\times$ 12& I & 1 & Conv. & 6 & VATT/GUFI\\

				& & & 2011 May 7 & 8.0 & 25 $\times$ 1& $i^{\prime}$ & ... & ...& 12 & VATT/4K\\

				& & & 2011 May 8 & 8.0 & 25 $\times$ 1& $i^{\prime}$ & ... & ... & 12 & VATT/4K\\ 
\hline \\
BRI 0021			& 3 & 28; 1.2 & 2009 Sept 14 & 4.0 & 5 $\times$ 12 & I & 1 & Conv. & 1 & VATT/GUFI\\

				& & & 2009 Sept 16 & 5.1 & 5 $\times$ 12& I & I & Conv. & 1 & VATT/GUFI\\
				
				& & & 2010 Nov 13 & 4.0 & 5 $\times$ 12& I & 1 & Conv.& 1 & VATT/GUFI\\

				& & & 2010 Nov 14 & 5.5 & 5 $\times$ 12 & I & 1 & Conv. & 1 & VATT/GUFI\\

				& & & 2010 Dec 2 & 5.1 & 5 $\times$ 12& I & 1 & Conv. & 1 & VATT/GUFI\\

				& & & 2010 Dec 3 & 4.5 & 5 $\times$ 12& I & 1 & Conv.& 1 & VATT/GUFI\\ 
\hline \\
2M J0036			& 2 & 10; 0.03& 2010 Dec 1 & 5.5 & 5 $\times$ 24 & I & 1 & Conv. & 6 & VATT/GUFI\\

				& & & 2010 Dec 13 & 5.0 & 5 $\times$ 24& I & 1 & Conv.& 6 & VATT/GUFI

\enddata

\tablecomments{Column (1) Campaign source. (2) The number of epochs over the course of the campaign. All epochs may contain multiple nights of observations, where these are not always sequential. (3) The total amount of hours on target in hours, and the total temporal baseline in years. (4) Observation dates for each target. (5) The length of each observation, as shown in the relevant figures in \S~\ref{results}. (6) The exposure time of each observation, as well as the binning factor used for final data points as shown in this paper. (7) The wave band used for a particular observation. (8) The readout rate used, in MHz. This column only applies to the GUFI mk.II system. (9) The amplifier used. Again, only applicable to GUFI mk.II. (10) The number of reference stars used for a given observation. We highlight that since the VATT 4K Imager provided a FOV of $\sim12.5^{\prime}\times12.5^{\prime}$, many more reference stars were available when compared to the smaller $\sim3^{\prime}\times3^{\prime}$ FOV of GUFI mk.II. Furthermore, there was one available star suitable for effective differential photometry in the case of BRI 0021. Although we could not confirm its stability against another non-varying star in the same field, we chose to use this based on the observations of \citet{martin01}, who confirmed it as a stable source during their photometric observations (indicated as reference star 1, in Figure 1 of their work). (11) Telescope and detector used.}\label{table_obs_details}

\end{deluxetable*}

\section{OPTICAL OBSERVATIONS}\label{optical_obs}

\subsection{GUFI mk.II - the Galway Ultra Fast Imager Photometer}\label{gufi}

\par The GUFI instrument was originally commissioned by astronomers in NUI Galway as an optical photometer capable of high-time resolution imaging \citep{sheehan08}. We modified the GUFI mk.II system (hereafter GUFI) to be compatible with the 1.83 m VATT on Mt. Graham, Arizona, where it is currently stationed as a visitor instrument. The system uses the Andor iXon DV887 EM-CCD camera, which has a CCD97 thinned back-illuminated sensor from e2v technologies, hosting $>$90\% quantum efficiency (QE) with a native 512 $\times$ 512 frame transfer sensor. It offers variable readout rates up to 10 MHz and can operate full-frame at 34 frames per second (fps) and up to 526 fps in a windowed configuration. The native field of view (FOV) of GUFI at the VATT Cassegrain focus is $\sim1.7^{\prime}$ $\times$ $1.7^{\prime}$ with a corresponding plate scale of 0.2$^{\prime\prime}$ pixel$^{-1}$. Focal reducer options for wider fields are limited by the short VATT back focal distance of 50.8 mm, but GUFI provides near-infrared (NIR) and visible-optimized focal reducers, offering a FOV of $\sim3^{\prime}$ $\times$ $3^{\prime}$ and a larger plate scale of 0.35$^{\prime\prime}$ pixel$^{-1}$. The VATT telescope offers the full range of Johnson and Sloan filter sets, as well as Vilnius interference filters, thus GUFI had an effective spectral sensitivity during this campaign of $\sim$3000 - 10000 \r{A} (based on the QE). The great advantages of GUFI for this study is its 100\% observing duty cycle (with a $\sim$2 ms readout rate), very low readout noise and high quantum efficiency.

\subsection{The VATT 4K Imager}\label{4k}

The VATT 4K CCD camera is the primary in-house photometer stationed at VATT. It houses a back-illuminated STA0500A CCD with a transfer sensor of 4064 $\times$ 4064 pixels, a native FOV of $\sim12.5^{\prime}$ $\times$ $12.5^{\prime}$ and a plate scale of 0.188$^{\prime\prime}$ pixel$^{-1}$. The standard readout rate for the camera is 50 seconds, however faster readout rates can be achieved based on the level of windowing applied to the frame. 

\subsection{The USNO Detectors}\label{usno}

Some observations as outlined in the relevant target details in \S~\ref{objects}, were obtained with the USNO 1.0 m and USNO 1.55 m telescopes. The new2k camera on the 1.0 m telescope has a FOV of $23.2^{\prime}$ $\times$ $23.2^{\prime}$ and a pixel scale of 0.68$^{\prime\prime}$ pixel$^{-1}$. We used the Tek2k camera on the 1.55 m, which has a corresponding FOV of $11.3^{\prime} \times 11.3^{\prime}$ with a pixel scale of 0.33$^{\prime\prime}$ pixel$^{-1}$.

\subsection{Observations and Data Reduction}\label{obs_reduc}

\begin{figure}[!t]
\epsscale{1.0}
\plotone{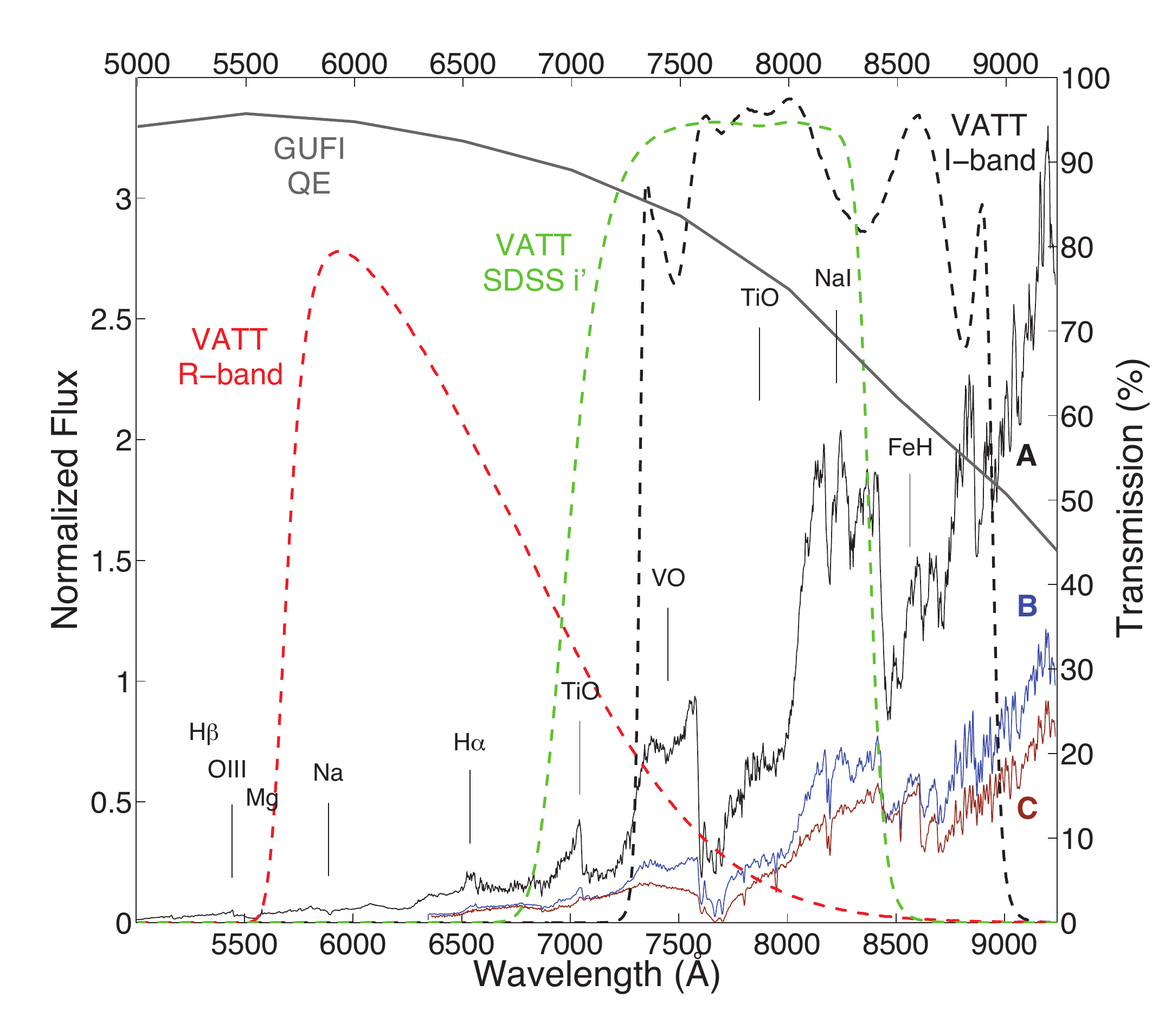}
\caption{Filter transmission curves with over-plotted spectra of an M8 dwarf (A), an L0.5 dwarf (B) and an L3.5 dwarf (C) - the spectral range which encompasses our study. The spectra have been normalized (y-axis, left) by the peak flux of the M8 dwarf spectra at 9200 \r{A}. The optical filters used in this study are shown by the dashed lines of wavelength (x-axes) vs. \% transmission (y-axis, right): VATT R-band ($\sim$5600 - 8800 \r{A}; red), Sloan $i^{\prime}$ ($\sim$6500 - 9500 \r{A}; green) and VATT I-band ($\sim$7200 - 9100 \r{A}; black). We also include the GUFI photometer's QE curve (solid grey line) to highlight transmission for the different wavebands.}\label{fig1}
\end{figure}

\par The observation campaigns were carried out between May 2006 - May 2011. We used the VATT R-Harris ($\sim$5600-8800 \r{A}) filter, the Sloan $i^{\prime}$ ($\sim$6500 - 9500 \r{A}) filter, the Johnson I-band filter ($\sim$7000 - 11000 \r{A}), and the VATT I-Arizona ($\sim$7200 - 9100 \r{A}) filter - for selected targets (Table~\ref{table_obs_details}). Transmission curves for each filter are shown in Figure~\ref{fig1}. The campaign encompassed observations to search for periodic variability of all radio detected dwarfs listed in Table~\ref{table_obs_details}, that were visible from the VATT observatory site (32$^{\circ}$42$^{\prime}$4.78$^{\prime\prime}$ 109$^{\circ}$53$^{\prime}$32.5$^{\prime\prime}$W). We also obtained data from the 1.52 m telescope, in Loiano, Bologna, Italy, as well as the 1.0 m and 1.55 m USNO telescopes, in Flagstaff, Arizona, as shown in Table~\ref{table_obs_details}. Figure~\ref{fig1} includes spectra of an M8.5, an L0.5 and an L3.5 dwarf, which covers the range of ultracool dwarf spectral types that our observations covered. Typical acquisition parameters are also summarized in Table~\ref{table_obs_details}.

\par Data reduction was carried out using the in-house GUFI L3 Pipeline \citep{sheehan08}. Standard data reduction techniques were employed where the data were bias subtracted using zero-integration frames and flat-fielded using twilight flat-fields. Twilight flat-fields for any given observation consisted of $>$100 median-combined dithered frames taken from a blank part of the sky. Frames were registered and summed in image space to increase the S/N, and differential photometry was carried out on all science data in order to achieve milli-magnitude photometric precision.

\par The FOVs of the GUFI, the VATT 4K and the USNO photometers, provide between 1-20 reference stars for a given field. Photometry for all reference stars was also obtained as a measure of their stability in order to ensure that variability was intrinsic to the target star. These stars were chosen on the basis of their stability, position, isolation, the properties of their seeing profiles, and comparable magnitudes and color to that of the target. Photometric apertures (in pixels) which provided the highest S/N for the target star were selected for aperture photometry; however aperture and sky annulus diameters varied from night to night depending on the average seeing conditions, which typically ranged from 0.7 to 1.6 arcseconds. Differential photometry was obtained by dividing the target flux by the mean flux of selected reference stars. Although changing seeing conditions can ultimately introduce photometric errors, for all observations we ensured that the photometric parameters remained constant for all stars - this allowed the same fraction of total flux to be observed in the aperture of each source.

\section{Assessing Periodic Variability}\label{periodic_variability}

In order to detect periodic variability and assess its significance, we used a variety of statistical tests as a means of assessing the validity of any detected periodic signals, and to calculate the associated errors. This assessment was carried out in order of the procedures below. These are well established techniques and so we only briefly explain each in the relevant sections - we refer the reader to the references therein for more in-depth discussions.

\subsection{Lomb-Scargle Periodogram}\label{LS}

\par The first method used for the detection of periodic signals was the calculation of the LS periodogram \citep{lomb76,scargle82}, a technique which is effective for unevenly spaced data. The LS periodogram uses the discrete Fourier transform (DFT), which provides power spectra that are analyzed for significant peaks - corresponding to possible periodic variability. In the case of an arbitrary (unevenly) sampled dataset, the LS periodogram is calculated by the following (where the power spectrum $P$, is a function of angular frequency $\omega=2\pi f>0$):

\begin{eqnarray}P(\omega)=\frac{1}{2\sigma^2_{var}}\frac{[\sum_{i}(h_{i}-\bar{h}) \cdot cos \cdot \omega (t_{i}-\tau )]^{2}}{\sum_{i}cos^{2} \cdot \omega (t_{i}-\tau )} + \nonumber\\ \frac{[\sum_{i}(h_{i}-\bar{h}) \cdot sin \cdot \omega (t_{i}-\tau )]^{2}}{\sum_{i}sin^{2} \cdot \omega (t_{i}-\tau )} \end{eqnarray}

where $\tau=tan(2 \cdot \omega \cdot t)=(\sum_{i}sin \cdot 2 \cdot \omega t_{i}/\sum_{i}cos \cdot 2 \cdot \omega t_{i})$, each consecutive data point is $h_{i}$, the mean of the data is $\bar{h}$ and the variance is $\sigma^{2}_{var}$.

\par In this work, we selected a range of peaks corresponding to possible periodic solutions as provided by the technique above. We inspected these solutions by phase connecting raw light curves to a given solution, and assessed their level of agreement in phase. We rule out solutions $>$0.25 out of phase. In addition, we over-plotted LS power spectra for different epochs, investigated which peaks were in greatest agreement, and then compared these to the strongest phase folded solutions. 


\begin{deluxetable*}{lccccccc}
\tablecaption{Peak to Peak Amplitude Variability and Photometric Error Analysis of Sample}

\tablehead{
\colhead{Source} & \colhead{Date of Obs.} & \colhead{Band} & \colhead{PtP$_{tar}$} &  \colhead{Phot. Error} &  \colhead{PtP$_{tar}$ Range} & \colhead{Mean $\sigma_{ref}$}\\
& (UT) & & (\%) & (\%) & (\%) & (\%)  \\ \\
 \colhead{(1)} &  \colhead{(2)} &  \colhead{(3)} &  \colhead{(4)} &  \colhead{(5)} & \colhead{(6)} & \colhead{(7)} 
}
\startdata
LP 349-25AB 	&2009 Sept 22 & I & 0.48 &  0.15& 0.44 - 1.42 (I); 1.96 (R) & 0.30 (I); 0.68 (R) \\

				&2009 Sept 26 & I & 1.42 & 0.21& &\\
				
				&2010 Oct 9 & I & 1.04& 0.21& & \\

				&2010 Oct 10 & I & 0.90& 0.22& & \\

				&2010 Oct 11 & I & 0.44& 0.18& & \\

				&2010 Oct 12 & I & 0.94& 0.21& &\\

				&2010 Oct 13 & I & 0.90& 0.15& &\\

				&2010 Oct 14& I & 0.58& 0.15& & \\

				&2010 Oct 15 & R & 1.96& 0.53 & &\\
				
				&2010 Nov 16 & I & 1.12& 0.23& & \\

				&2010 Nov 27 & I & 0.92& 0.15& & \\
\hline \\
2M J0746AB		&2009 Jan 25 & I & 0.40& 0.21& 0.40 - 1.52 (I) & 0.36 (I)\\

				&2009 Jan 26 & I & 0.98& 0.28& &\\
				
				&2009 Jan 28 & I & 0.78& 0.24& &\\

				&2010 Feb 19 & I & 1.26& 0.27& &\\

				&2010 Feb 20 & I & 1.32& 0.30& &\\

				&2010 Nov 13 & I & 1.18& 0.31& & \\

				&2010 Nov 14 & I & 1.04 &  0.33& &\\

				&2010 Dec 2 & I & 0.68& 0.25& & \\

				&2010 Dec 12 & I & 1.38& 0.29  & &\\

				&2010 Dec 13 & I & 1.52& 0.32& &\\

				&2010 Dec 14 & I & 0.96& 0.34& &\\
\hline \\

LSR J1835		& 2006 Jul 17 & I & 1.08& 0.12& 1.02 - 1.46 (I); 1.62 (R) & 0.33 (I); 0.68 (R) \\

				&2006 Jul 20 & I & 1.02& 0.13& & \\
				
				&2006 Sept 22 & I & 1.46&0.43& & \\

				&2006 Sept 24 & R & 1.62& 1.20 & & \\

				&2009 Jun 11 & I & 1.24& 0.12 & &\\

				&2009 Jun 13 & I & 1.34& 0.16 & & \\

				&2009 Jun 16 & I & 1.32& 0.12 & & \\

				&2009 Jun 30 & I & 1.36& 0.18 & & \\
\hline \\

TVLM 513 		&2006 May 21 & I & 0.82& 0.42& 0.56 - 1.20 (I); 0.92 - 0.96 ($i^{\prime}$) & 0.34 (I); 0.36 ($i^{\prime}$)\\ 

				&2008 Jun 17 & I & 0.66&0.53& &\\
				
				&2009 Jun 12 & I & 0.56& 0.30& &\\

				&2009 Jun 13 & I & 0.72& 0.23& &\\

				&2009 Jun 16 & I & 1.14& 0.25& & \\

				&2011 Feb 18 & I & 1.20& 0.32& &\\

				&2011 Feb 25 & I & 0.70& 0.32 & & \\

				&2011 Apr 12 & I & 0.76& 0.31& & \\

				&2011 May 7 & $i^{\prime}$ & 0.96 & 0.27& &\\

				&2011 May 8 & $i^{\prime}$ & 0.92& 0.26 & &\\
\hline \\
			
BRI 0021	    & 2009 Sept 14 & I & 1.10 & 0.33& 0.52 - 1.58 (I) & 0.37 (I)\\

		        &2009 Sept 16 & I & 0.90 &  0.32& &\\
				
				&2010 Nov 13 & I & 0.72 & 0.32& &\\

				&2010 Nov 14 & I & 1.58 & 0.31& &\\

				&2010 Dec 2 & I & 0.68 & 0.32& &\\

				&2010 Dec 3 & I & 0.52& 0.35& &\\

\hline \\

2M J0036		&2010 Dec 1 & I & 2.20& 0.82& 1.98 - 2.20 (I) & 1.0 (I)\\

				&2010 Dec 13 & I & 1.98& 1.11& & 

\enddata

\tablecomments{Column (1) Campaign source. (2) Date of observation in UT. (3) Waveband used. (4) Peak to peak (PtP) amplitude variability as measured by the $\chi^{2}$ test. (5) Mean photometric error per data point for a given night as calculated by the \textit{iraf.phot} routines. This is outlined in \S~\ref{error}. (6) Peak to peak amplitude variability range of target light curves, shown in I-band or Sloan $i^{\prime}$, and in R-band, for selected targets. Both R-band results are from single observations. (7) Standard deviation of non-variable reference star light curve in R-band, Sloan $i^{\prime}$ and I-band (mean standard deviation of all reference stars used in each case).}\label{table_amp_var}

\end{deluxetable*}

\subsection{Phase Dispersion Minimization}\label{PDM}

\par We also investigated the PDM technique as outlined by \citet{stellingwerf78}, as a second statistical tool. \citet{stellingwerf78} describes the PDM method as a least squares fit (LSF) approach where a fit is calculated by using the mean curve of the data, controlled by the mean of each bin (which can be specified in the algorithm), and the period that produces the least datapoint scatter, or `PDM theta statistic' ($\Theta$), about this computed mean, is the most likely solution.

\par The PDM technique phase folds selected light curves to a range of periods, and their significance is calculated. It is useful for data sets with large gaps, and furthermore, it is insensitive to the light curve's shape and therefore makes no assumptions with regard to the morphology. The routine also includes a Monte-Carlo test, used for assessing the statistical significance of the detected $\Theta$ minima. It computes this by randomizing the data point order, which removes the signal component. We repeated this for 10$^{5}$ trials in order to cover a significant distribution of $\Theta$ values due to noise\footnote[9]{We cite \citet{stellingwerf78} for the PDM routines, but refer to his latest work at http://www.stellingwerf.com/.}. Similar to the LS technique above, it is possible for many periodic solutions to present themselves due to aliasing - a consequence of gaps in the data. We take the minimum $\Theta$ from the PDM analysis, and compare it to the highest peak in the power spectra of the LS.

\subsection{Amplitude Variability Analysis}\label{amp_var}

\par We established the peak to peak amplitude variability of the target light curves by means of sinusoid fitting and the $\chi^{2}$ technique, where the phase and amplitude of a sinusoidal function were varied, and then the $\chi^{2}$ minimization was performed. We took this amplitude (which is a peak to peak (PtP) measure of the change of relative flux) as PtP$_{tar}$. This is a weighted assessment and so does not treat each data point equally; the error in each point is utilized in the calculation of the best fit amplitude and the error in the amplitude. 

\par The corresponding reference star variability was found via the standard deviation of its light curve ($\sigma_{ref}$). We plotted each reference star flux against all others to ensure that each chosen selected reference star was non-variable. Although variability can statistically be detected if the standard deviation is only fractionally larger than the error in the light curve's relative magnitude, the periodic variability detected in our target data is categorically present in each epoch, where the variability is clearly above the standard deviation of the reference star relative flux. Furthermore, different sets/combinations of reference stars were used as a `sanity check' to confirm that the signal was indeed intrinsic to the target star.

\subsection{Photometric Error Estimation}\label{error}

\par The photometric error analysis was calculated via the \textit{iraf.phot}\footnote[10]{Image Reduction and Analysis Facility - http://iraf.noao.edu/.} routines in all target and reference star light curves. An estimation of the error in the relative magnitude ($\delta m_{\star}$) of the target star was found as follows:

\begin{equation}(\delta m_{\star})^{2} = (\delta_{target})^{2} + \left(\frac{1}{MF_{i}}\right)^{2} \sum_{n}^{M} F^{2}_{n} (\delta m_{n})^{2} \end{equation}  

where \textit{M} is the number of reference stars, \textit{F$_{i}$} is the mean flux of the reference stars, \textit{F$_{n}$} is the flux of the \textit{n$_{th}$} reference star and $\delta m_{n}$ is the magnitude error in the \textit{n$_{th}$} reference star. This error in magnitude was then converted to an error in flux. We show these error bars on each data point in each light curve. This method takes both formal and informal errors such as flat-fielding and residual fringing (\S~\ref{fringing}) into account - which are difficult to assess in separate cases. 

\par In addition to the formal and informal errors, we also identify detector response at non-linear regimes as a source of potential error. We avoid such non-linear effects by keeping exposure times low enough to maintain levels to no greater than 75\% of pixel saturation. After taking these effects into account, we move to calculating the period uncertainty.

 \begin{deluxetable*}{lccccccc}
\tabletypesize{\scriptsize}
\tablecaption{Confirmed Optical Periodic Variability in Radio Detected Ultracool Dwarf Sample}
\tablewidth{0pt}

\tablehead{
\colhead{Parameter} & \colhead{LP 349-25B} & \colhead{2M J0746A} & \colhead{LSR J1835} & \colhead{TVLM 513} & \colhead{BRI 0021} & \colhead{2M J0036}  
}
\startdata
(1) Period (hrs) ....................		& 1.86 $\pm$ 0.02 & 3.32 $\pm$ 0.15 & 2.845 $\pm$ 0.003 & 1.95958 $\pm$ 0.00005& ? ($\sim$5) & $\sim$3.0 $\pm$ 0.7 \\

(2) LS Period (hrs)	...............	& 1.86 &3.32 & 2.845 & 1.95958&...& 2.5\\

(3) PDM Period (hrs) ...........		& 1.86 & 3.32 &2.844& 1.95959& ...& 2.5\\

(4) References .......................		& 1 & 1, 2 & 1  & 1, 3 & 1 & 1, 3

\enddata

\tablecomments{Row (1) Period of rotation and associated error as calculated in \S~\ref{periodic_variability}. (2) Lomb-Scargle Periodogram periods: the quoted periods are those which were determined to be the most likely solution based on the correlation of the highest peaks in all periodograms (all data combined and individual epochs). (3) Phase Dispersion Minimization periods: the PDM periods shown here represent the lowest $\Theta$ statistic calculated by the PDM routines, as is shown in \S~\ref{results}.\\ \\ References. --- (1) This work. (2) \citet{harding13}. (3) \citet{lane07}: TVLM 513 originally published as $\sim$1.96 hours, 2M J0036 published as $\sim$3 hours.}\label{table_results}

\end{deluxetable*}

\subsection{Fringing}\label{fringing}

\par Fringing is an optical effect or disturbance in the thinned-substrate of back-illuminated CCDs and is present as a result of OH spectral emission in the atmosphere. Fringing interferes at red/NIR wavelengths and since the CCD's substrate becomes transparent at these wavelengths, any waveband that approaches the NIR is more susceptible to these fringing effects. It varies as a function of amplitude, but not position. Since the amplitude variations expected in these ultracool dwarf targets are of the order of milli-magnitudes, it is important to remove these additive effects if the amplitude variations due to fringing are potentially greater than the target star differential light curves. The standard procedure for this correction includes the creation of a fringing template from well sampled median-combined deep sky frames containing only the fringing pattern, normalizing this template to each individual frame's sky background level and then subtracting it. We obtained dithered sky frames for all Sloan $i^{\prime}$ and I-band observations to allow for fringe removal if necessary. We also took dome flat-fields which contain none of these atmospheric effects, in addition to twilight flat-fields. We conducted tests to investigate the effect of this artifact on each consecutive data set, and if the amplitude of the fringing pattern was varying at a greater level than that of the mean sky background, it was removed.

\subsection{Phase Connecting \& Period Uncertainty Estimation}\label{period_uncertainty}

\par We achieve an accurate enough period of rotation for the M9 dwarf TVLM 513 to phase connect its $\sim$5 year baseline. We could not phase connect any other target, and thus the procedure outlined here applies to TVLM 513 only. Standard phase connection techniques were employed whereby the period accuracy increased as epochs were successfully phase connected, enabling an assessment of the correlation of the peak of each phase solution. This allowed us to combine data from two different epochs, if the period from a single epoch could be calculated with sufficient accuracy, such that the rotational phase of the second epoch was unambiguous - in this work we define this threshold to be $\delta \phi < 0.25$.

\par In order to assess the period error for all other targets, we overplotted the LS power spectra period range with a Gaussian profile, and calculated the FWHM. In this way, we estimate 1$\sigma$ errors on the period uncertainty ($\delta P$) for these targets. Since the $FWHM=2\sqrt{2ln2} \; \sigma=2.35482 \sigma$, $\delta P$ is therefore defined as:

\begin{equation}\delta P=\frac{FWHM}{2.35482} \end{equation}

\par We find that the uncertainty range calculated for each target for the best-fit period of rotation, allowed other possible solutions within this range to be phased together within epochs. The $\chi^{2}$ test outlined in the previous section also provided a measure of the period error per given fit. Other authors have also established various means of assessing the error in the frequency of a signal, e.g. \citet{schwartz91,akerlof94}. These techniques can largely rely on data uniformly sampled in time. Thus, similar to the $\chi^{2}$ fitting, they were effective in calculating an error for a single observation, but not for unevenly spaced baselines.

\begin{figure*}[!m]
\epsscale{1.0}
\plotone{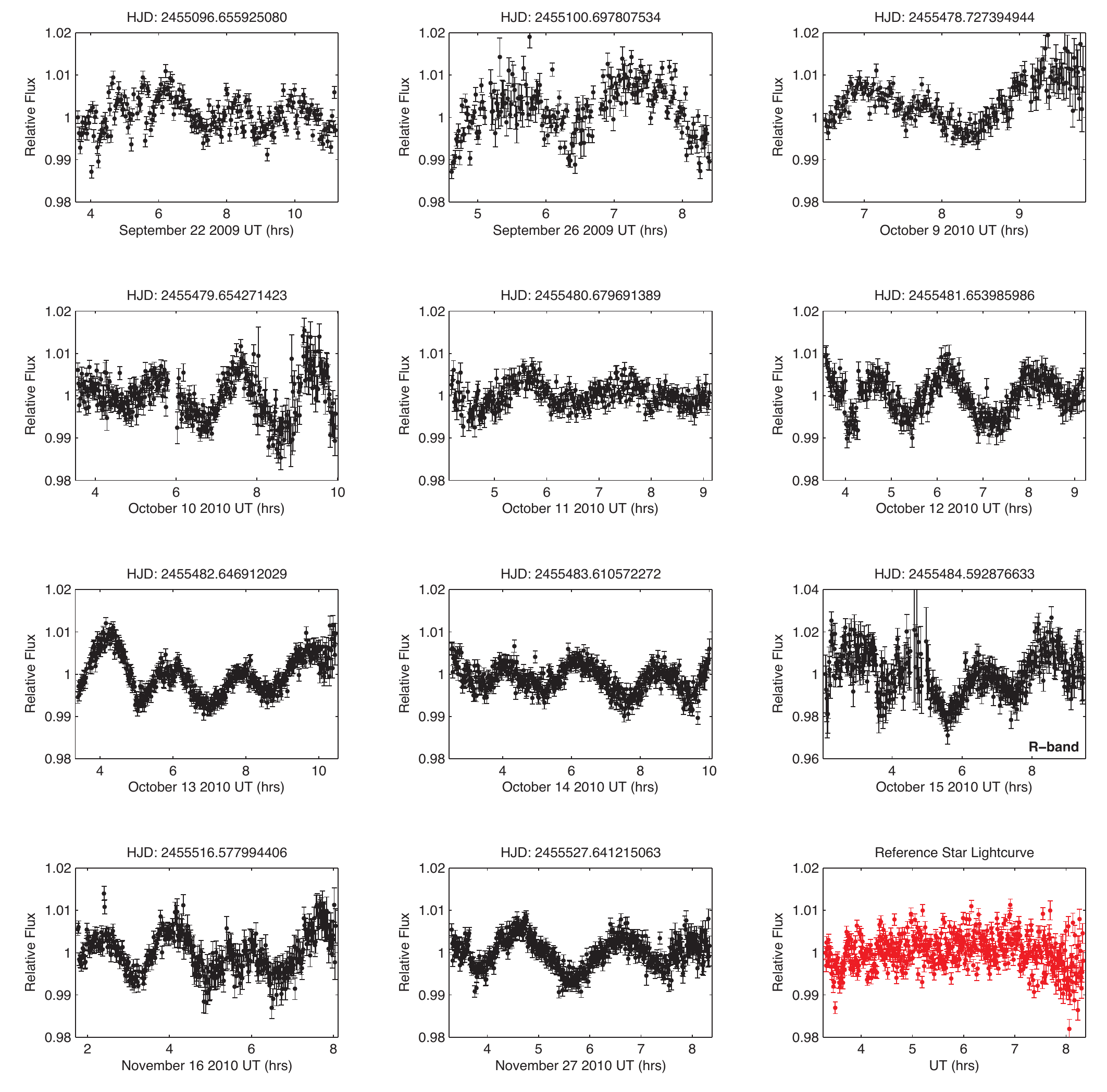}
\caption{LP 349-25: Photometric light curves showing Relative Flux (y-axis) vs. UT dates and times (x-axis). The HJD time above each figure denotes the start-point of each observation. It is important to note that the x-axis range is not the same for each plot, since observations were of different lengths. All data in this figure was taken in VATT I-band ($\sim$7200 - 9100 \r{A}), with the exception of October 15 2010 UT which was taken in VATT R-band ($\sim$5600-8800 \r{A}) - this is marked on the relevant light curve. Note the difference in scale on the y-axis for the R-band labeled plot. We detect periodic variability that shows a persistent period of 1.86 $\pm$ 0.02 hours over $\sim$1.2 years of observations. These data exhibit changes in amplitude in I-band during consecutive nights (e.g. Oct 10, 11, 13: $\sim$0.44 - 1.42\%), as well as some aperiodic variations observed during some observations (e.g. Oct 9). The R-band light curve exhibits larger peak to peak amplitude variations of 1.96\%; The second R-band peak in the signal was an interval of poor weather conditions (thin cloud) shown clearly by an increase in the photometric error measurements. The September 2009 epoch was also subject to poor weather conditions (intermittent cloud \& thin cloud throughout), and was therefore binned by a factor of 2 compared to the other data. Photometric error bars are applied as outlined in \S~\ref{error}. [\textit{bottom right}] - we selected a reference star at random, and plotted its raw flux against the mean raw flux of all other reference stars used in the field. This is used as an example of reference star stability compared to target variability. We note that this light curve is an example of one night only, however we used the same reference stars for all epochs in a given band. The mean reference star variability for all reference stars used in this campaign is shown in Table~\ref{table_amp_var}.}\label{fig2}
\end{figure*}

\begin{figure*}[!t]
\epsscale{1.0}
\plotone{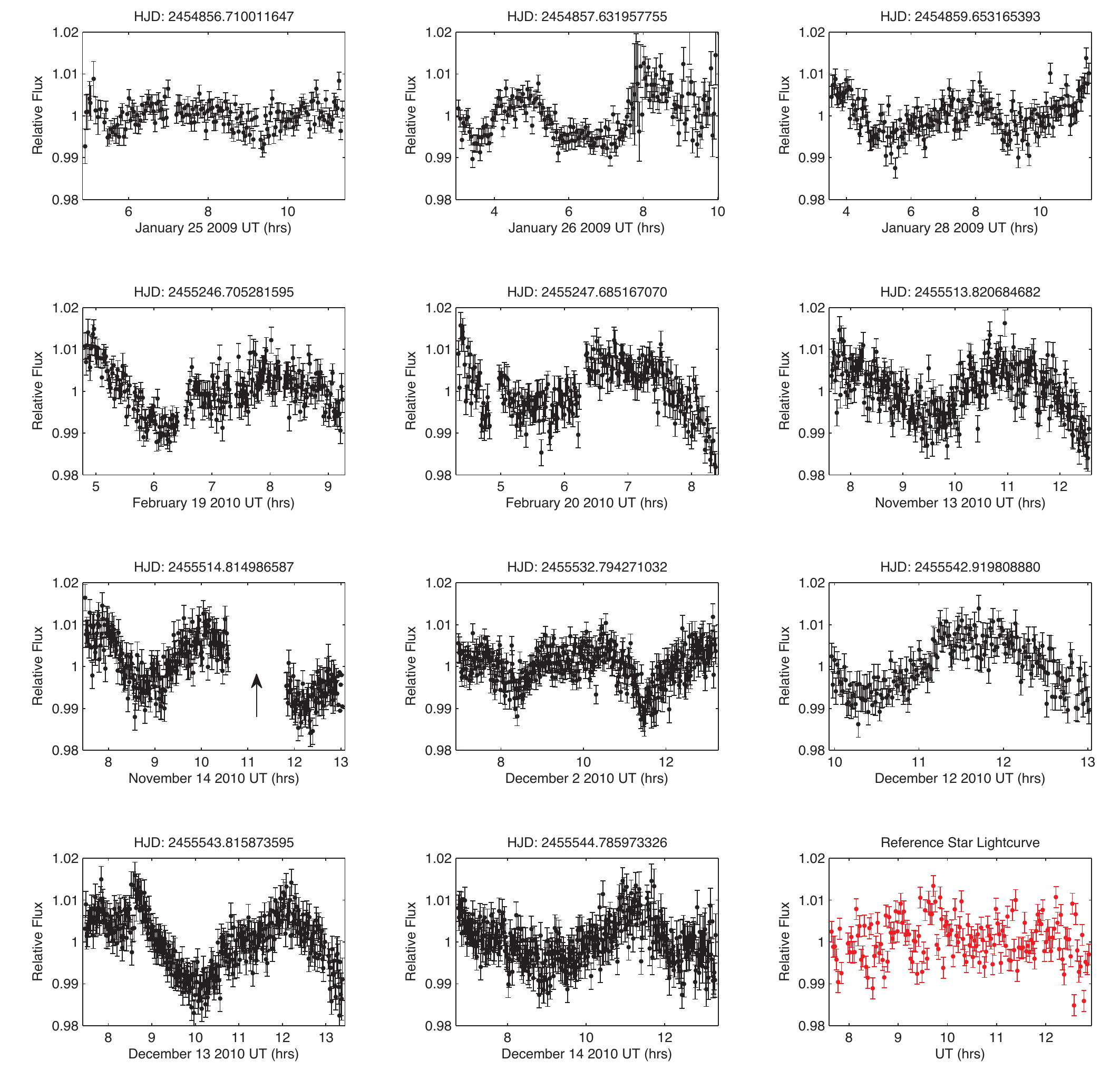}
\caption{2MASS J0746+2000: Photometric light curves first reported by \citet{harding13}, and included here for completeness to investigate emission morphology and behavior. Again, UT dates and times are marked on each light curve's x-axis along with HJD time above each figure (start-point of each observation). These data were taken in VATT I-band ($\sim$7200 - 9100 \r{A}) over an $\sim$2 year baseline. We report periodic variability for one component of the binary, with a period of 3.32 $\pm$ 0.15 hours. The peak to peak amplitude variations throughout the observations varies from $\sim$0.40 - 1.52\%. We note that January 25 \& 26 2009 were taken during deteriorating weather conditions (thin cloud and high winds) and were therefore binned by a factor of 2 compared to other data. The arrow marked on the November 14 2010 light curve points to an interval of complete cloud cover, therefore these data were removed. Photometric error bars are applied to each data point as before. [\textit{bottom right}] - as before, an example reference star light curve to illustrate the stability of the chosen reference stars as compared to the target star variability. The mean reference star variability for all reference stars used in this campaign for 2M J0746 is shown in Table~\ref{table_amp_var}.}\label{fig3}
\end{figure*}

\begin{figure*}[!t]
\epsscale{1.0}
\plotone{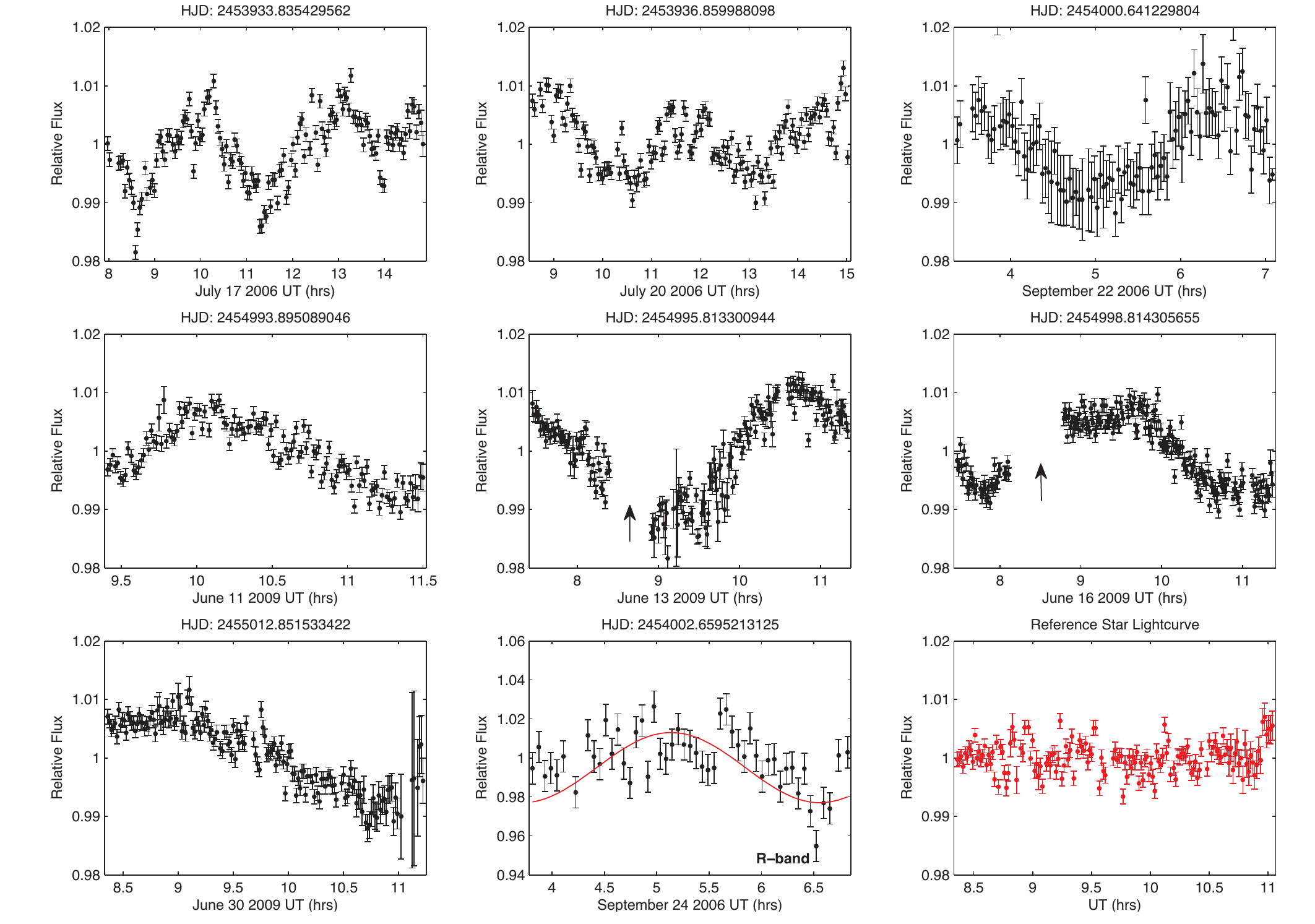}
\caption{LSR J1835+3259: We report a photometric period of rotation of 2.845 $\pm$ 0.003 hours at I-band wavelengths ($\sim$7000 - 11000 \r{A}) using the GUFI photometer. These data were taken over a $\sim$3 year baseline, where the 2006 July epoch was taken as test data for the GUFI mk.I system. We also observed the dwarf in R-band ($\sim$5600-8800 \r{A}) using the USNO on September 24 \& 25 2006 UT. The seeing on both nights was very poor however. Here we show a binned data set, marked with an R-band label, from September 24 2006 UT. We overplot a model sinusoidal fit (red) to a period of 2.845 hours. The period of rotation of 2.845 $\pm$ 0.003 hours matches the periodic pulses reported by \citet{hallinan08}, who also attributed this periodicity to the dwarf's rotation. The arrows shown in June 13 \& June 16 mark data gaps due to this object's passing too close to the zenith for the telescope's Alt-Az tracking. Once again we show a reference star light curve (bottom right) to illustrate the variability of the target with respect to a non-variable source. Although we have a $\sim$3 year baseline, we do not achieve an accurate enough period to phase connect the 2006 and 2009 epochs.}\label{fig4}
\end{figure*}

\begin{figure*}[!t]
\epsscale{1.0}
\plotone{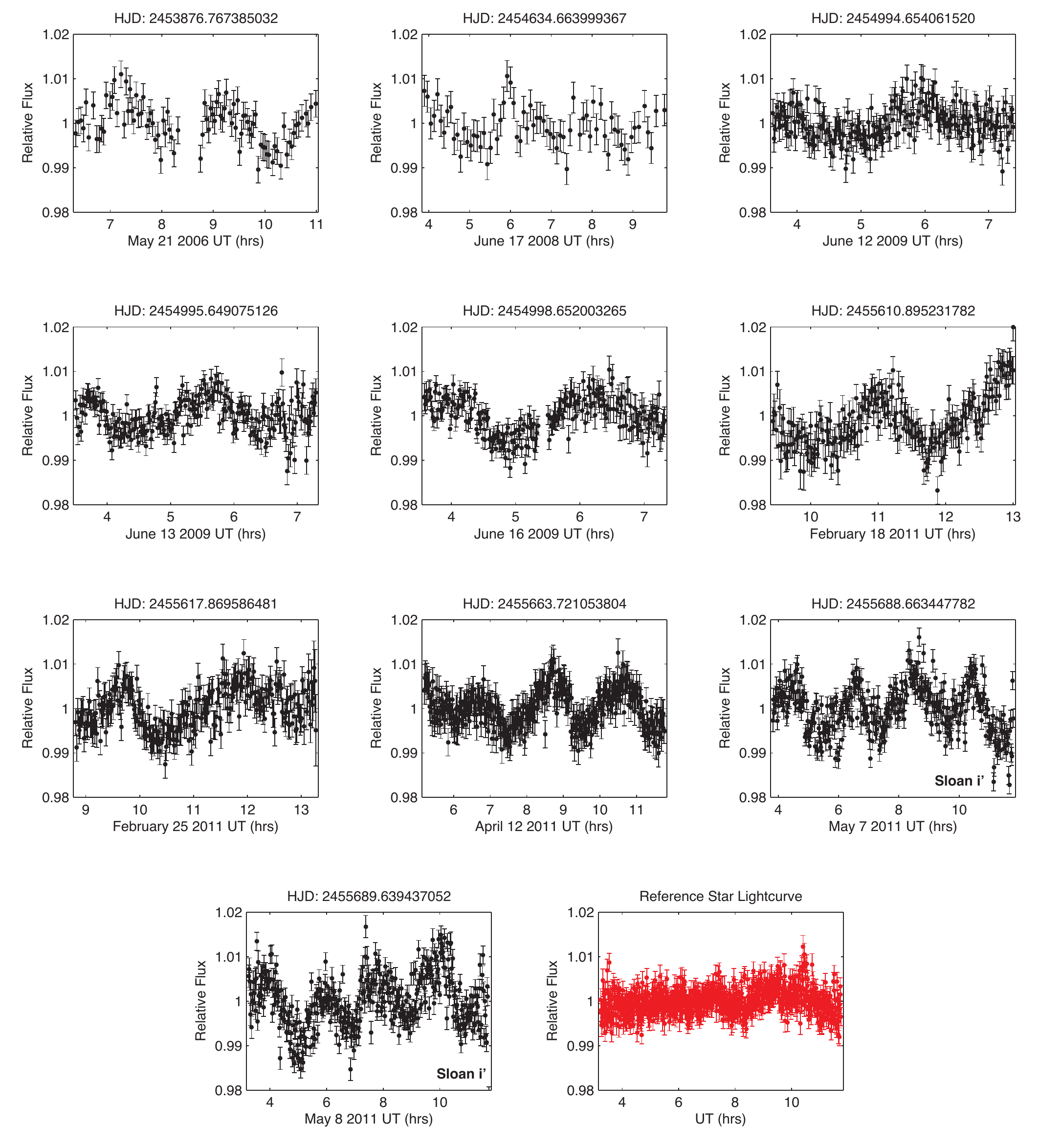}
\caption{TVLM 513-46546: We obtained $\sim$53 hours of data, over a $\sim$5 year baseline for TVLM 513. Our data shows an extremely stable period of 1.95958 $\pm$ 0.00005 hours, which we phase connect over this baseline. The data shown here was taken in I-band ($\sim$7000 - 11000 \r{A}) and Sloan $i^{\prime}$ ($\sim$6500-9500 \r{A}), which is marked on the relevant light curves (May 7 \& May 8 2011). This confirmed period further constrains the work of \citet{lane07} who found a photometric period of $\sim$1.96 hours, also in I-band. As in the case of LSR J1835, this periodicity is consistent with the observations of \citet{hallinan06,hallinan07}, who report periodic radio pulses of $\sim$1.96 hours for TVLM 513. In this work, we investigate the stability of the light curve phase and amplitude, and find the phase to be stable throughout each data set, where changes in amplitude are present (0.56 - 1.20\% in I-band and 0.92 - 0.96\% in Sloan $i^{\prime}$). We discuss this further in the following section. As always, a randomly selected reference star light curve is included (bottom right).}\label{fig5}
\end{figure*}

\begin{figure*}[!t]
\epsscale{1.0}
\plotone{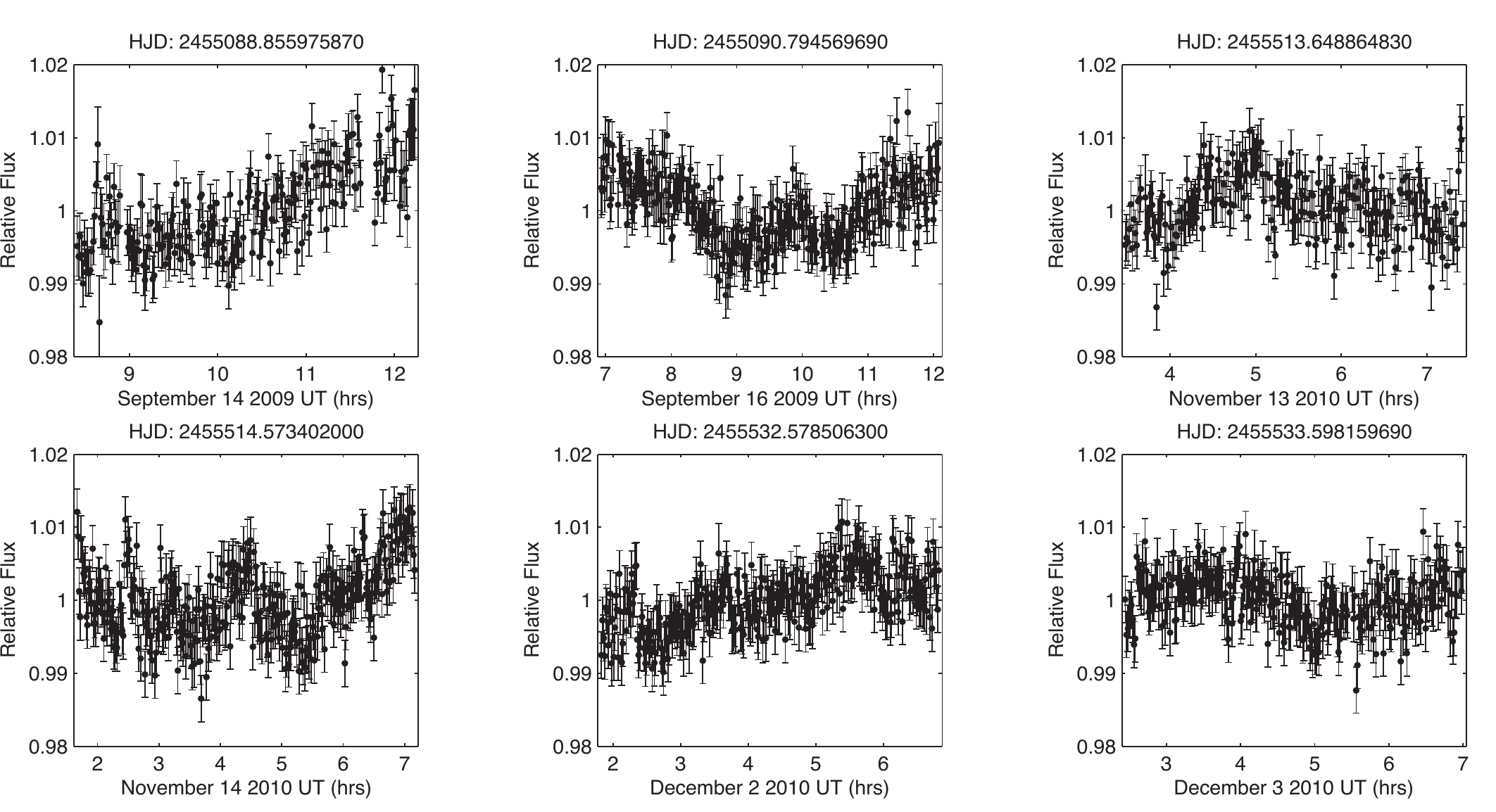}
\caption{BRI 0021-0214: We observed BRI 0021 for a total of 6 nights, over 3 epochs. Previous studies by \citet{martin01} found evidence for variability, with possible periods of $\sim$4.8 hours and $\sim$20 hours. The 3$^{\prime}\times3^{\prime}$ FOV of GUFI only allowed for one suitable reference star however (00$^{h}$ 24$^{m}$ 23$^{s}$.735,-01$^{\circ}$ 59$^{\prime}$ 06.27$^{\prime\prime}$). We selected this on the basis of its stability which was assessed by \citet{martin01}. We report possible periodic variability with peak to peak amplitude variations of 0.52 - 1.58\%. Although the periodograms show favorable evidence for a period of $\sim$5 hours, we take this only as a tentative estimate due to the behavior observed in the light curves above; i.e. we could not constrain one likely solution for all epochs without imposing large errors. However, it is worth noting that a period of $\sim$5 hours is in conflict with current \textit{v} sin \textit{i} estimates for the system, and would infer that the stellar radius has been underestimated. Further observations, with more field stars, and larger temporal coverage are needed to effectively assess the photometric behavior of this object.}\label{fig6}
\end{figure*}

\begin{figure}[!t]
\epsscale{0.68}
\plotone{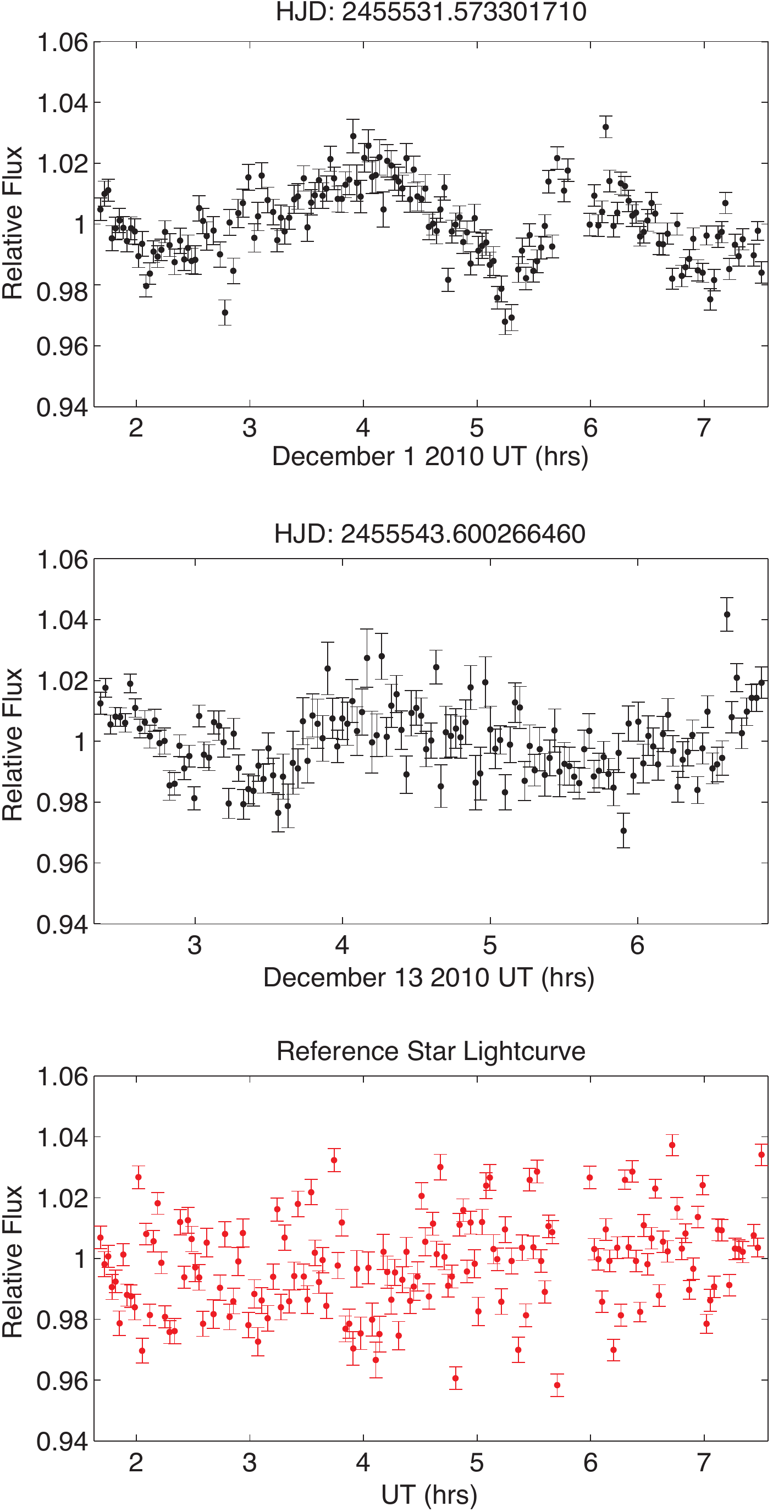}
\caption{2MASS J0036+18: We confirm a period of 3.0 $\pm$ 0.7 hours for 2M J0036. Unfortunately, both nights of observation were subject to poor weather conditions (heavy cloud). Nevertheless, our range of periods are in agreement with the observations of \citet{lane07}, who detect a $\sim$3 hour period for this source in the Johnson I-band. \citet{berger05,hallinan08} showed this dwarf to be radio pulsing with a period of 3.08 $\pm$ 0.05 hours. We note that the light curves above were binned to 2 minute frames in order to increase the S/N.}\label{fig7}
\end{figure}

\begin{figure*}[!m]
\epsscale{1.0}
\plotone{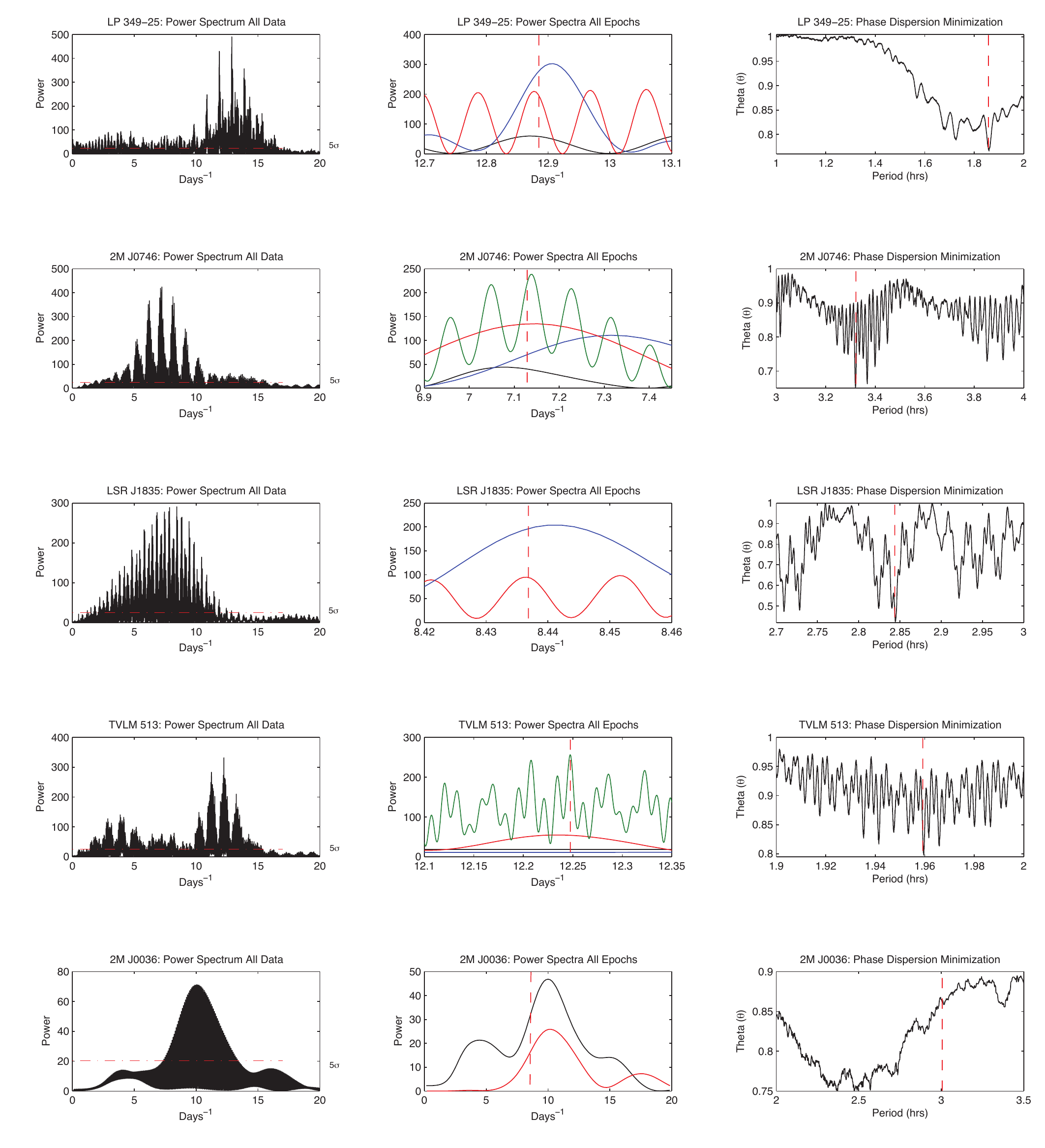}
\caption{[\textit{Column LEFT \& column MIDDLE}] Lomb-Scargle Periodograms for all periodically detected sources. The \textit{left column} shows periodograms (power spectra) for each target for all epochs of observations. We include a red dashed-doted horizontal line on each plot which represents a 5$\sigma$ false-alarm probability of the peaks as determined by the Lomb-Scargle algorithm in each case. We also we point out multiple power spectra peaks centered around the highest peaks that correspond to the reported rotation periods (\textit{left column}). These peaks are present as a result of spectral leakage, which is due to large gaps in the data between consecutive epochs. Each figure in the \textit{middle column} once again shows a periodogram plot for individual epochs (over-plotted) to illustrate period correlation between each. [\textit{LP 349-25}] Black - Sept 2009; Blue - Oct 2010; Red - November 2010. [\textit{2M J0746}] Black - Jan 2009; Blue - Feb 2010; Red - November 2010; Green - Dec 2010. [\textit{LSR J1835}] Red - 2006 data; Blue - 2009 data. [\textit{TVLM 513}] Black - May 2006; Blue - June 2008; Red - June 2009; Green - 2011 data. [\textit{2M J0036}] Black - Dec 1 2010; Red - Dec 13 2010. We note for TVLM 513 in particular, the amplitude of the May 2006 (black) and June 2008 (blue) power spectra is much lower than the other epochs, and thus appears flat on this plot. The x-axis (Days$^{-1}$) of each figure is scaled to the approx. period range as calculated by our uncertainty technique, with the exception of 2M J0036 where we show the full range of assessed values due to poorer temporal coverage. We also include a red vertical dashed line corresponding to the established period of rotation in this work. [\textit{Column RIGHT}] - Phase Dispersion Minimization plots for each target, showing a plot of period against the `Theta' ($\theta$) statistic. This statistic was determined based on 10$^{5}$ Monte-Carlo simulations which randomize the data points and test whether the result at any given $\Theta$ level could be as a result of noise. The most significant periods are marked with a red dashed line on each figure. In the case of 2M J0036, we mark the period of $\sim$3 hours as detected by \citet{lane07} and confirmed in this work. The variability analysis was more difficult for this target due to poor photometric conditions.}\label{fig8}
\end{figure*}

\section{RESULTS}\label{results}

\subsection{General Results}

\par We report periodic variability for five of the six radio detected dwarfs in the sample, shown in Table~\ref{table_results}. The properties of this periodicity is generally consistent for all dwarf spectral types, where we detect periodic sinusoidal variability over time scales of years. Our assessment of the peak to peak amplitude variations for each target are shown in Table~\ref{table_amp_var}. All dwarfs exhibit changes in amplitude throughout the campaign, which we discuss in \S~\ref{properties}.

\par In the following subsections, we outline general results and variability analysis of each target, as well as light curve and photometric properties. All confirmed periods in these data were detected to significance values exceeding 5$\sigma$. The target results are shown through Figures~\ref{fig2} - \ref{fig7}, and the variability analysis for each is shown in Figure~\ref{fig8}. We discuss the possibilities for the cause of this periodic variability in \S~\ref{properties}.

\subsection{Binary Dwarfs}\label{binaries}

\subsubsection{LP 349-25}

\par We detect the binary as a periodically varying source in VATT R-band and I-band, which we report as the first detected optical variability of this system. The primary period of 1.86 $\pm$ 0.02 hours is present in each band and varying with a PtP$_{tar}$ range of 0.44 - 1.42\% in I-band, and 1.96\% in R-band (single observation), as shown in Table~\ref{table_amp_var} and Figure~\ref{fig2}. The LS periodogram and PDM statistical analysis is shown at the end of the section in Figure~\ref{fig8}. Mean $\sigma_{ref}$ were calculated to be $\sim$0.30\% and $\sim$0.68\% in I-band and R-band, respectively. We see larger $\sigma_{ref}$ in R-band due to intermittently poor seeing. It is difficult to assess the amplitude ratios between each band, since the amplitude level in the I-band is varying at different levels during observations (Table~\ref{table_amp_var}). Furthermore, we did not obtain simultaneous R-band and I-band data.

\par Despite the consistency of the primary periodic component throughout the observations, we observe some aperiodic variations in addition to significant variations in amplitude during some I-band observations (e.g. Figure~\ref{fig2}: Oct 10, 11 \& 13 2010). We do not image each component of the binary as a single point source in these observations, therefore the detected sinusoidal periodicity in our data is due to the combined flux of both binary members. We observe unusual behavior for some of the October 2010 epoch, where the periodic signal appears to move in and out of phase \textit{during} single observations of $\sim$8 hours; we give examples of this in \S~\ref{lp349_phase_stab}. 

\par Finally, the radii estimates of \citet{dupuy10} and individual rotation velocity measurements of \citet{konopacky12} infer maximum rotation periods of $\sim$2.65 hours and $\sim$1.67 hours for each component, respectively. Therefore, we have a tentative case to argue in favor of LP 349-25B as the periodically varying source in R- and I-band wavelengths. However, the radii estimates of \citet{konopacky10} are at odds with those derived in this work as well as the estimates of \cite{dupuy10}. This modeling, and the association of the 1.86 hour period with LP 349-25B, are discussed later in \S~\ref{lp349_coplanar}.

\subsubsection{2MASSW J0746425+200032}

\par The periodic variability of 2M J0746AB has recently been discussed by \citet{harding13}, who use this rotation period to infer the coplanarity of the spin axis and orbital plane. We include a discussion of the variability here again for completeness. Although we do not resolve each component of the binary as a point source, we report optical periodic modulation of 3.32 $\pm$ 0.15 hours from 2M J0746A, with peak to peak amplitude variability of PtP$_{tar}$ $\sim$ 0.40 - 1.52\% in VATT I-band (Figure~\ref{fig3}), and a mean reference star standard deviation of $\sigma_{ref}$ $\sim$0.36\%.

\par It appears that this optical periodic variability originates from the \textit{other} component to that producing the radio emission - reported by \citet{berger09} where the binary exhibited periodic bursts of radio emission of 2.07 $\pm$ 0.002 hours. The estimated radii of $\sim$0.99 $\pm$ 0.03 $R_{J}$ and $\sim$0.96 $\pm$ 0.02 $R_{J}$ \citep{harding13}, in addition to the well established \textit{v} sin \textit{i} measurements \citep{konopacky12}, infer maximum rotation periods for 2M J0746A and 2M J0746B of $\sim$4.22 hours and $\sim$2.38 hours, respectively. Therefore, the period of 3.32 $\pm$ 0.15 hours likely emanates from 2M J0746A, whereas \citet{berger09} found emission from the secondary - 2M J0746B. This optical periodicity is categorically present in all epochs as shown in Figure~\ref{fig3}, and thus is that of the \textit{slower} rotating binary dwarf.

\vspace{1cm}

\subsection{Single Dwarf Systems}\label{single_sys}

\subsubsection{LSR J1835+3259}

\par We determined a photometric period of 2.845 $\pm$ 0.003 hours in VATT I-band, consistent with the VLA radio observations of \citet{hallinan08}, who report periodic pulses of 2.84 $\pm$ 0.01 hours. This optical period is newly reported in this work, which was conducted between July 2006 and June 2009 with the GUFI mk.I and mk.II systems (Figure~\ref{fig4}). We also obtained R-band data from the 1.55 m USNO telescope, and detected periodicity of $\sim$2.84 hours. The weather for this observation was very poor; however it appears that LSR J1835 has larger R-band peak to peak amplitude variability than I-band - similar to LP 349-25. These data exhibit long-term stable periodic sinusoidal variability with a PtP$_{tar}$ range of 1.02 - 1.46\% in I-band and 1.62\% in R-band. The standard deviation of the selected reference stars in each band were $\sigma_{ref}$ $\sim$ 0.33\% and $\sim$0.68\%, respectively. Furthermore, the calculated period supports the rotational velocity estimate of \textit{v} sin \textit{i} $\sim$50 $\pm$ 5 km s$^{-1}$ \citep{berger08}, and radius estimate of $\geq$0.117 $\pm$ 0.012 $R_{\odot}$ \citep{hallinan08}, which implies a high inclination angle of $\sim$90$^{\circ}$ for the system. These data also appear to be in phase based on this period of 2.845 $\pm$ 0.003 hours during constituent epochs. However we do not achieve a high enough period accuracy in order to phase connect the $\sim$3 year temporal baseline. We show the statistical analysis for this target in Figure~\ref{fig8}. An example of reference star stability is also shown in red in Figure~\ref{fig4}, bottom right.

\subsubsection{TVLM 513-46546}

\par We confirm periodic variability of 1.95958 $\pm$ 0.00005 hours, with a peak to peak amplitude variability range of PtP$_{tar}$ $\sim$0.56 - 1.20\% in VATT I-band and  PtP$_{tar}$ $\sim$0.92 - 0.96\% in Sloan $i^{\prime}$. The morphology of the light curves are generally consistent for both wavebands throughout the campaign, with a mean $\sigma_{ref}$ of I: $\sim$0.34\% and $i^{\prime}$:$\sim$0.36\%. The larger peak to peak amplitude variations for some observations are shown in Table~\ref{table_amp_var}. This period once again supports previous studies from \citet{hallinan06, hallinan07}, \citet{lane07}, \citet{berger08} and \citet{littlefair08}, and a clear indication that the photometric I-band periodic variability appears to be stable over time-scales of up to 5 years in this case. It is also consistent with the radius, \textit{v} sin \textit{i} and inclination angle estimates outlined in \citet{hallinan08}. The calculated PtP$_{tar}$ in I-band is lower than the reported peak to peak amplitude variability of \citet{lane07}. However, the $i^{\prime}$ variability is much higher than that observed by \citet{littlefair08}, who detect PtP$_{tar}$ of only $\sim$0.15\% in their data. Light curves from each of the six epochs are shown in Figure~\ref{fig5} and the LS periodogram and PDM analysis is shown in Figure~\ref{fig8}. In \S~\ref{phase_stab}, we show phase connected light curves over the 5 year baseline in order to investigate the target's phase stability - this study directly investigates the positional stability of the stellar feature responsible for the periodicity. By phase connecting the total baseline of TVLM 513, we were able to establish a period to a much greater accuracy than other targets where phase connection was not possible, due to limited phase coverage.

\subsubsection{BRI 0021-0214}

\par We report possible photometric VATT I-band periodic variability with PtP$_{tar}$ of $\sim$0.52 - 1.58\%, and $\sigma_{ref}$ of $\sim$0.37\%, which is shown in Figure~\ref{fig6}. We note that due to $\sim3^{\prime}\times3^{\prime}$ FOV of GUFI, there was only one suitable reference star used for differential photometry. This star was selected as a suitable candidate on the basis of its observed stability compared to the target star, during the I-band observations of BRI 0021 by \citet{martin01}. They identify possible periodicity of $\sim$4.8 hours and $\sim$20 hours, respectively. We do not have sufficient temporal coverage to effectively assess the presence of a $\sim$20 hour period. Although there is evidence in our statistical analysis of periods between 4 - 7 hours, we do not sample the rotational phase of the object enough to confirm a solution. Since we only have one reference star as a comparison source (00$^{h}$ 24$^{m}$ 23$^{s}$.735,-01$^{\circ}$ 59$^{\prime}$ 06.27$^{\prime\prime}$), its stability cannot be independently assessed in this case. Interestingly, the possible solutions of $\sim$4 - 7 hours are in violation with the current \textit{v} sin \textit{i} estimates of $\sim$34 km s$^{-1}$ found by \citet{mohanty03} - which indicate a maximum period for this system of $\sim$3.59 hours. This indicates that the radius of the dwarf could be underestimated if a periodic signal $>$3.59 hours is present. Further (larger FOV) observations, with greater temporal coverage on a given night are needed to constrain and qualitatively confirm this result.

\subsubsection{2MASS J00361617+1821104}

\par We confirm sinusoidal periodic variability of 3.0 $\pm$ 0.7 hours with PtP$_{tar}$ of 1.98 - 2.20\% in the optical VATT I-band. Although these data were obtained under extremely poor seeing conditions on both nights of observation, the range of periods within the calculated error matches the $\sim$3 hour periodicity found by the photometric measurements of \citet{lane07} and the radio measurements of \citet{berger05} and \citet{hallinan08}. We note that the observed PtP$_{tar}$ is larger than that of other I-band data in this work. We show the differential light curves in Figure~\ref{fig7}, and the analysis of these in Figure~\ref{fig8}.

\section{Discussion}\label{properties}

\subsection{Source of the Periodicity: the Optical-Radio Correlation?}

\par A large number of surveys have been carried out to search for evidence of optical variability in ultracool dwarfs. In this work, we consider only late M to early-to-mid L dwarfs. Beyond this point, it is clear that the variability has predominantly been associated with dust-related effects \citep{artigau09,radigan12}. To date, 182 ultracool dwarfs in this spectral range ($\geq$M6 - L5) have been studied for optical variability, where there has only been $\sim$30 - 40\% of confirmed variability \citep[and references therein]{tinney99,bailerjones99,jonesmundt01,gelino02,clarke02b,koen03,koen04,koen05,rockenfeller06,maiti07,koen12}. In many cases, these studies have yielded low variability detection rates, or tentative detections with low significance \citep{koen03,enoch03,koen04,maiti07,goldman08}. Others have found more promising statistically significant detection rates where the variability was clearly detected above the noise-floor \citep{jonesmundt01,gelino02,rockenfeller06}. Considering the spectral range in our survey ($\geq$M7.5 - L3.5) compared to this same range in the above studies of late-M and early-to-mid L dwarfs, less than 5\% of objects studied have \textit{confirmed} periodic variability consistent with the rotation period \citep{clarke02a,koen03,koen06,rockenfeller06,lane07,koen11}. 

\par Our study has confirmed periodic optical variability for five out of six radio active dwarfs, with a tentative detection of similar behavior in the sixth; the latter case limited by poor sampling of the rotational phase of the object. However, a direct comparison to a large fraction of the above work will show that our sensitivities are much higher for detecting periodic variability in these objects (see \S~\ref{optical_obs} for GUFI specs). Throughout this campaign, we have consistently achieved photometric precisions of $<$0.5\% (and as low as 0.15\% for some observations) as shown in Table~\ref{table_amp_var}, as well as sampling many rotation periods per object, per night. By contrast, the above studies have typically achieved photometric precisions of $\geq$1.0\% (with some as low as $\sim$0.5\%), and in many cases the rotational phase has been poorly covered. Furthermore, large-survey data sets only produced a few data points per hour in order to contemporaneously monitor a large number of objects. Thus, a combination of high photometric precision, well sampled rotational phase coverage and high cadence data sets, are perhaps crucial in effectively detecting periodic variability from ultracool dwarfs on these time scales. In particular, we highlight that two of our sample, 2M J0746AB and 2M J0036, were included in surveys mentioned above \citep{clarke02b,maiti07}. In both cases, variability was detected, albeit with insufficient phase coverage to recover the periodic signal detected in our work. 

\par However, we cite \citet{jonesmundt01} and \citet{rockenfeller06} as reliable comparison studies (with similar sensitivities to the periodic variability reported in this work) that searched for variability (including periodic variability) in a sample of dwarfs that were not pre-selected as radio detected. We select objects in these papers between $\geq$M7.5 - L3.5 only, in order to satisfy a direct comparison to the objects in this work. In the case of \citet{jonesmundt01}, they observed 21 M and L type dwarfs, where 15 of these occupy the spectral range we define above. Similarly, \citet{rockenfeller06} cover a sample of 19 M dwarfs, where 6 of these are M7.5 - M9. Crucially, their work provided detection limits of $\sim$0.5 - 5\% in the peak to peak amplitude variations of variable targets. In addition to this, the methods of \citet{jonesmundt01} were sensitive to periods $>$1 hour, and \citet{rockenfeller06} for $\sim$0.5 - 12 hours. Therefore, based on the peak to peak amplitudes detected in our work, their studies both had the capability of detecting the presence of periodic variability in their sample. However, although \citet{jonesmundt01} report evidence of some periodic signals, they only report tentative detections from some targets. Similarly, \citet{rockenfeller06} detect periodic optical variability from only 1 source out of 6 - the M9 dwarf 2MASSW J1707183+643933. During these observations, they also report a large flare event \citep{rockenfeller06b}, and as a result argue that the presence of magnetic activity is expected. \citet{gizis00} reported H$\alpha$ emission for the same M9 dwarf with an equivalent width of 9.8 \r{A}, further supporting the possible presence of magnetic activity.

\par The presence of consistent periodic variability in five of six radio detected ultracool dwarfs demonstrates that the correlation between optical and radio periodic variability is significant and thus the presence of magnetic activity is also significant when compared to the above studies. We therefore have a case to highlight an expected presence of consistent periodic optical variability in radio detected sources, due to the presence of strong magnetic fields (kG) with radio activity. We note further that all of our target sample are rapid rotators, with high \textit{v} sin \textit{i} values ($>$15 km s$^{-1}$). This is perhaps an additional bias in our data, whereby rapid rotators could be easier detected than slowly rotating sources. However, an expanded sample of non radio-active dwarfs that are also rapid rotators, is required to quantify this further.

\par Previous studies have argued that magnetic spots, (e.g. \citet{rockenfeller06,lane07}) or dust (e.g. \citet{jonesmundt01,littlefair08}) were responsible for similar detected periodicities in ultracool dwarfs. One possible means of distinguishing between various mechanisms is to compare simultaneous multi-band photometry to synthetic atmospheric models \citep{allard01}. Importantly, in order to carry out such analyses, simultaneous observations are needed due to the inherent variability in the amplitude of the optical variability. The studies of \citet{rockenfeller06} and \citet{littlefair08}, for example, yielded cases in favor of both cool magnetic spots and the presence of atmospheric dust, respectively. These results were based on the ratios of the peak to peak amplitude variations at different photometric wavebands. Despite detecting larger peak to peak amplitude variations in R-band vs. I-band for two of our target sample (see Table~\ref{table_amp_var}), we did not obtain simultaneous photometry and therefore cannot apply these models at this point. However, such a high detection rate of significant periodic variability in our sample of radio active dwarfs, implies a correlation with radio activity and thus some kind of magnetic phenomenon. The nature of this optical variability will be addressed in an upcoming paper focused on spectro-photometric observations of such targets (Hallinan et al., in prep).

\begin{figure*}[!t]
\epsscale{1.0}
\plotone{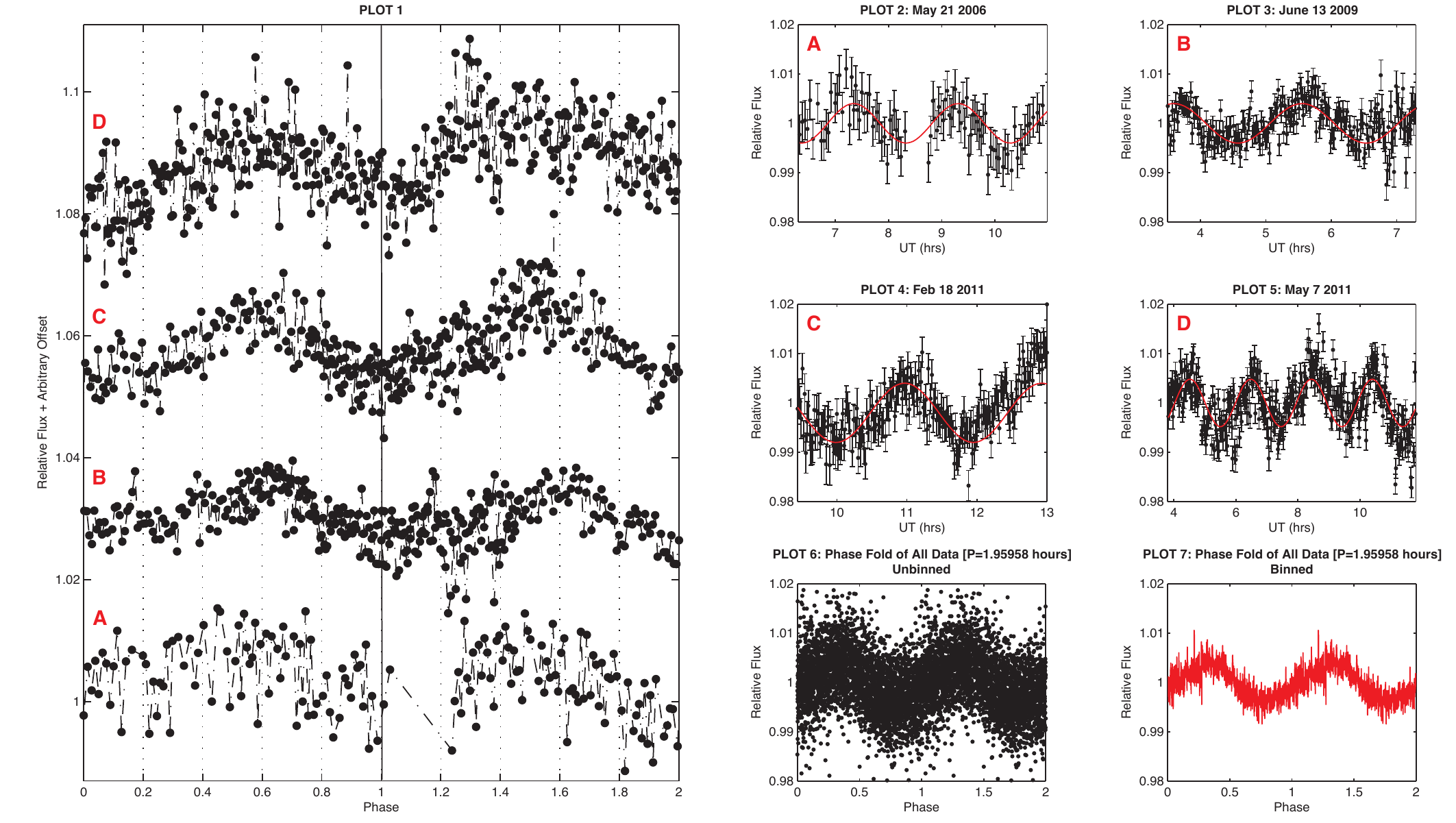}
\caption{\textit{[PLOT 1]}: This figure illustrates the phase stability of the periodic variability of TVLM 513 over a $\sim$5 year baseline. These raw light curves, labeled with red letters A - D (bottom - top), were selected at random from four of the observation epochs (May 2006 - May 2011). This level of agreement is consistent for all light curves in the sample. In each case, the time stamps were phase folded to the period of 1.95958 hours. \textit{[PLOTS 2 - 5]}: To show this agreement further, the light curves A, B, C \& D in PLOT 1 correspond to PLOTS 2, 3, 4 \& 5, respectively. Each light curve contains an overplotted model sinusoidal signal (red), with a period of 1.95958 hours, which was applied to the full 2006 - 2011 dataset, where we set values between individual observations and epochs to zero. It is clear that this dwarf exhibits highly correlated behavior in terms of phase over this baseline, and furthermore that the stellar feature responsible must be equally as stable (spatially) during these observations. \textit{[PLOT 6 \& PLOT 7]}: We phase fold the entire data set (2006 - 2011, containing $\sim$3,500 data points) to the detected period of 1.95958 hours. The black phase folded light curve in PLOT 6 is raw and has no binning or scaling. The red phase folded light curve in PLOT 7, once again of all data, has been binned by a factor of 10.}\label{fig9}
\end{figure*}

\begin{figure*}[!t]
\epsscale{1.0}
\plotone{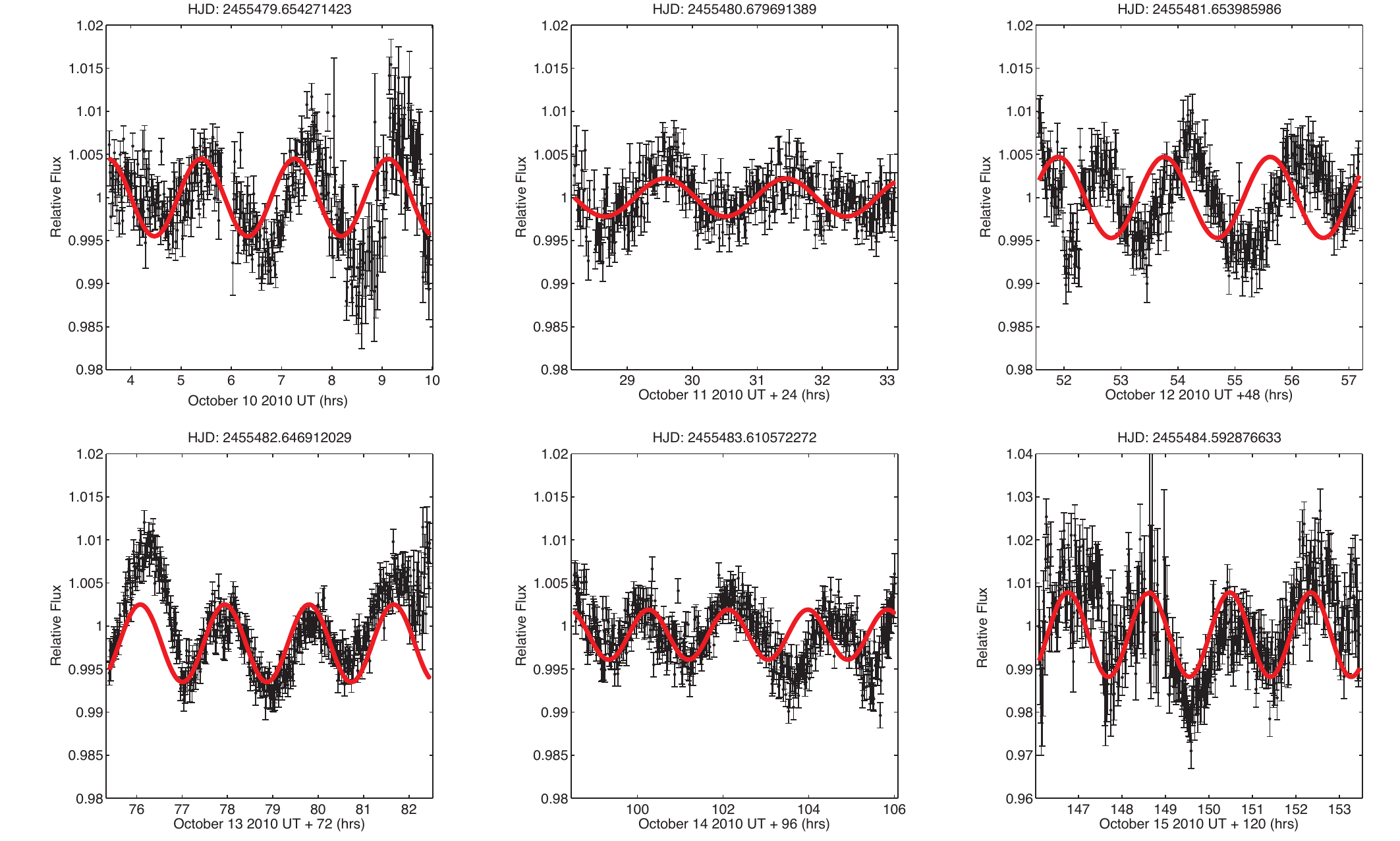}
\caption{LP 349-25: We show the behavior of LP 349-25 periodic variability from the October 2010 epoch (October 10 - October 15 UT). We have also overplotted a model sinusoidal fit (red) of $P = 1.86$ hours, the primary periodic component detected in our data. Amplitude values were taken from Table~\ref{table_amp_var} for each night, but most importantly, we use a fixed phase for the sinusoidal model for all observations here. The model appears to move in and out of phase during observations. For example, the fit is clearly in phase at the beginning of October 10, but \textit{as the amplitude in the light curve gets larger}, the phase begins to move out ($\sim$7 UT). This same effect is seen for October 13 and 14. By contrast, the signal is in phase for October 11 but out of phase for October 12. This behavior is possible evidence of a dynamical environment in the source region of the optical variability. Alternatively, the superposition of two variable sources could cause changing amplitudes and phase. We cite TVLM 513 (Figure~\ref{fig9}) as an example of a source exhibiting consistent phase stability for an established period.}\label{fig10}
\end{figure*}

\subsection{The Phase Stability of TVLM 513-46546}\label{phase_stab}

\par A number of ultracool dwarfs have shown periodic behavior over a number of observations (e.g. \citet{berger05,hallinan06,hallinan07,lane07,littlefair08,hallinan08}); here we use multi-epoch observations of one dwarf, TVLM 513, to investigate whether this periodicity is long term and stable in phase, and whether this modulation evolves morphologically over these time scales. We achieve an accurate enough period of rotation of 1.95958 hours for the dwarf via phase connection of the 2006 - 2011 epochs, with an associated error in the period of 0.00005 hours, thereby allowing us to assess its modulated behavior over the $\sim$5 year campaign. We find long-term, periodic variability that is stable in phase as shown in Figure~\ref{fig9}, where we overplot a model sinusoidal signal (red) over the entire baseline. Phase folded light curves of individual observations are shown in \textit{PLOT 1} of Figure~\ref{fig9}, once again highlighting this agreement, and similarly in \textit{PLOTS 6 \& 7} we show phase folded light curves of all datasets.

\par Such a high degree of correlation suggests a spatially-stable surface feature that does not appear to move by a significant amount over this baseline. \citet{donati06} \& \citet{morin10} have shown that large-scale magnetic fields for fully convective objects are stable on year-long time scales. \citet{hallinan06,hallinan07,hallinan08} have confirmed the presence of stable kG magnetic fields for TVLM 513, consistent with a common magnetic field-related origin for the periodic radio and optical variability for ultracool dwarfs, as discussed in the previous section.

\par While stable in phase, the peak to peak amplitude is variable during this campaign from $\sim$0.56 - 1.20\% in VATT I-band (see Table~\ref{table_amp_var}). Since the phase is stable, a change in amplitude suggests that the intensity of the feature responsible is changing on these levels, or that it may be changing in size. \citet{littlefair08} observed peak to peak variations in Sloan $i^{\prime}$ of 0.15\%. Here we report much larger Sloan $i^{\prime}$ peak to peak variability of 0.92 - 0.96\% - further evidence of variable peak to peak amplitudes when compared to other studies. This is intriguing when compared to the previous radio activity discussed by \citet{hallinan06,hallinan07}, who reported highly variable signals from TVLM 513. Specifically, they detected bursts of periodic radio emission that varied greatly between epochs, also indicating changes in emission intensities. Whether the optical emission here is directly related to radio variability will be conclusively determined when multiple epochs of radio data are obtained, and phase connected, over the same time scales as this work.

\subsection{The Radio \& Optical Emission at Odds from 2MASS J0746425+200032AB?}

\par In this work, we have demonstrated evidence of a correlation between the optical and radio variabilities in ultracool dwarfs. We therefore briefly consider why we detect optical periodic variability from the non-radio detected binary component of the 2M J0746AB system. 

\par According to model-derived temperature estimates of \citet{konopacky10}, the effective temperature of 2M J0746A ($T_{eff}$ $\sim$ 2205 $\pm$ 50 K) is higher than its counterpart ($T_{eff}$ $\sim$ 2060 $\pm$ 70 K). As previously discussed, our photometry contains the combined flux of both stars - perhaps the contrast ratios of stellar photosphere vs. feature are much greater for 2M J0746A as a result. If the optical and radio emission are linked as we put forward as a possibility, why did \citet{berger09} not also observe some evidence of radio emission from 2M J0746A? 

\par The primary could be pulsing at radio frequencies, but undetectable due to the inclination angle of the system. However, \citet{harding13} find that the 2M J0746AB rotation axes are orthogonally aligned to the system orbital plane. This established alignment geometry could support detectable beaming from both stars. However, this is contingent upon the magnetic field alignment of each star being equal with respect to their rotation axes. Misaligned magnetic field axes could mean that the radio emission from 2M J0746A is being beamed away from observer. Alternatively, unlike 2M J0746B, it is also possible that 2M J0746A does not exhibit beamed ECM emission at all, but perhaps only small levels of quiescent radio emission that has not yet been detected by previous studies of the system. Speculating further about the intricacies of the system's radio emission and the associated beaming geometry is outside the scope of this work.

\par Some aperiodic variability is also present for some observations which could be due to the contribution from a weaker secondary signal. The Lomb-Scargle periodogram analysis in this work should extract both photometric signals if they are both present and strong enough, and our data shows strong evidence of variability of the non-radio emitting component. Resolved photometry would be an interesting confirmation if the radio-active source is also optically variable.

\subsection{The Unusual Behavior of LP 349-25}\label{lp349_phase_stab}

\par In this section we discuss the behavior of the light curves of the binary LP 349-25AB. As outlined in \S~\ref{binaries}, we observe significant changes in amplitude in I-band (refer to Table~\ref{table_amp_var}), as well as changes in phase during single observations. Due to the close separation of the binary members, the photometric aperture used enclosed the combined flux of both components. Therefore the presence of two periodically varying sources in these data and thus the superposition of these waves is one possible explanation for the varying amplitude we observe here. However, aperiodic variability of a single periodic source could also cause this behavior. This is an obvious distinction and one that we discuss below.

\par We first consider the possibility of the presence of two periodically varying sources by subtracting the main 1.86 hour period out of the raw data. We did this by generating a sinusoidal model wave function with a period of 1.86 hours. We then iterated through a range of amplitude and phase values, and performed a LSF fit to the raw data from the October 2010 epoch. We chose this set of data because we had contiguous observation nights from October 9 - October 15 2010 UT, as shown in Figure~\ref{fig10}. The best solution which fitted the raw data parameters was subtracted out. Lomb-Scargle periodogram analysis was run on the remaining data points, which searched for residual periodic signatures. We observed no obvious evidence in the periodogram of any second significant source. As a follow-up, we modeled the superposition of two sinusoidal sources by setting a period of 1.86 hours for one source, varying the other period, as well as the amplitude and the phase of both waves, and performed a LSF to our data - as outlined in \S~\ref{amp_var}. These fits did not yield strong evidence of another source based on the best LSF solutions. The lack of evidence in the periodograms, as well as the inability to clearly detect an underlying source in the residual data after subtracting the main 1.86 hour period out, does not support the obvious presence of another period.

\par Nevertheless, the varying component of amplitude and phase remains in these data, as shown in Figure~\ref{fig10}. In this plot, we show raw light curves from the October 2010 epoch (Oct 10 - Oct 15 UT) with a model sinusoidal wave overplotted (in red). The established period of 1.86 hours was used, and corresponding amplitudes from Table~\ref{table_amp_var} were adopted for each light curve. We use a fixed phase for all nights. As we observe the model wave for each observation, we can see that the wave is in phase for some nights (e.g. Oct 11, Oct 13 and Oct 15). By contrast, the signal appears to have moved out of phase for Oct 12. We can also see, for Oct 10 and Oct 14 for example, that the model is largely in phase for the first half of each observation (although upon closer inspection there is some evidence of trailing and leading peaks and troughs), but then moves partially \textit{out of phase} as the amplitude of the signal increases - we also note changes in light curve morphology for these sections. 

\par This behavior could be characteristic of a high-dynamic environment in these regions, where the source of the variability is evolving on these time scales. Perhaps a magnetic feature is not stationary on the stellar photosphere, or alternatively a combination of features could be effecting light curve morphology. Moreover, if these features were undergoing changes in size or temperature, this could also have an effect on the sinusoidal shape. We can not rule out the possibility of another source - perhaps a more robust modeling technique than those used here is required to identify the presence of another period. Obtaining a contiguous time series of LP 349-25 over many periods of rotation, would allow us to more effectively investigate whether these morphological changes are evolving in a systematic and repeatable manner.

\subsection{Spin-Orbit Alignment of LP 349-25AB}\label{lp349_coplanar}

\par The detected rotation period from LP 349-25B in this work provides an important parameter in assessing the orbital coplanarity of the system, as well as the associated implications for binary formation theory in the VLM binary regime. Recent work by \citet{harding13} has demonstrated spin-orbit alignment for the VLM binary 2M J0746AB - the first such observational result in this mass range. Their work showed that the spin axes inclinations of both components of the system were aligned to within 10 degrees of the orbital plane. Such an alignment signals that solar-type binary formation mechanisms, such as core fragmentation, disk fragmentation or competitive accretion, may extend into the realm of brown dwarfs. Although the alignment of one system could not be used to distinguish between the various formation theories, investigating such alignments in other VLM systems provides an insight into where the above formation pathways may dominate. Here we applied the same approach as outlined in \citet{harding13} to assess the orbital properties LP 349-25AB.

\subsubsection{Estimating age and mass}

\par We used the evolutionary models of \citet{chabrier00} to estimate the age and mass (and later the radius) of each binary component. These parameters were constrained by adopting the established total system mass of 0.121 $\pm$ 0.009 $M_{\odot}$, as well as the photometric \textit{J} \textit{H} \textit{K} measurements and bolometric luminosity measurements of \citet{konopacky10}. In addition, previous spectroscopic investigations yielded no lithium in the dwarf's spectrum, e.g. \citet{bouy04}. We used these parameters to identify a range of ages where lithium was absent, and next interpolated over a range of masses by comparing the correlation between the  \textit{J H K} colors and bolometric luminosities of \citet{konopacky10}, and those of the \citet{chabrier00} models. Furthermore, by assuming each component was coeval, the sum of the component masses could not exceed the measured total system mass of 0.121 $\pm$ 0.009 $M_{\odot}$. 

\par We find an age consistent with \citet{dupuy10} of $\sim$140 Myr, with masses of $\sim$0.06 $M_{\odot}$ and $\sim$0.05 $M_{\odot}$ for LP 349-25A and LP 349-25B, respectively. However, \textit{lithium is present} in this range. \citet{dupuy10} suggested that perhaps the absence of lithium in the binary spectrum was due to flux domination from the primary member, and given the predicted mass of LP 349-25B in their work, the Li$_{I}$ doublet is expected since LP 349-25B potentially lies below the theoretically predicted lithium depletion point at $\approx$0.055 - 0.065 $M_{\odot}$. The only ages (where Li=0) that are in mild agreement suggest that the system has a total mass that far exceeds 0.121 $\pm$ 0.009 $M_{\odot}$. Lithium however may not be a robust indicator of age. For example, \citet{baraffe10} point out that episodic accretion can cause lithium to be depleted at younger ages, despite its expected presence based on evolutionary models. Another possibility might also be that the total system mass has been under-estimated, which would place LP 349-25AB at an older age in the models of \citet{chabrier00}, consequently supporting the observed absence of lithium.

\begin{figure}
\begin{center}
\includegraphics[scale=0.4]{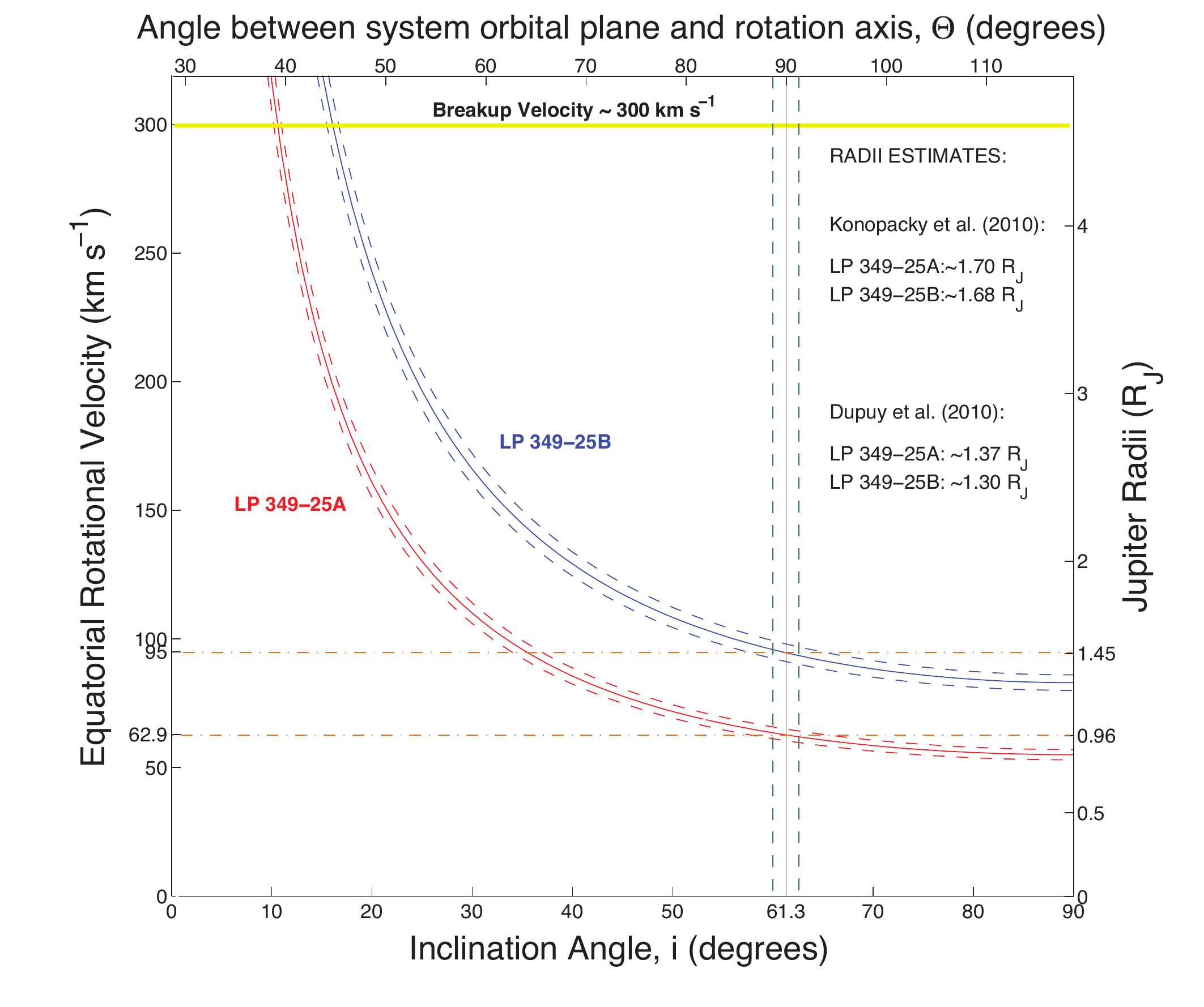}
\caption{Here we plot the \textit{v} sin \textit{i} of LP 349-25A (red) and LP 349-25B (blue) of \citet{konopacky12}. The dashed red and blue lines correspond to the error in this measurement. This figure investigates the radii estimates of \citet{konopacky10} \& \citet{dupuy10}, and whether the binary member's equatorial axes are coplanar with the system's orbital plane \citep{hale94}. We place one explicit constraint here: the presence of a rotation period of 1.86 $\pm$ 0.02 hours for one or other of the components. We illustrate this by aligning the measured system inclination angle of 61.3 $\pm$ 1.5 degrees, i, x-axis bottom) at 90 degrees to the equatorial axes (x-axis, $\Theta$, top); as shown by the green vertical line and the associated dashed error lines. \citet{konopacky12} report equatorial velocities of $\sim$62 km s$^{-1}$ and $\sim$95 km s$^{-1}$ for LP 349-25A and B, respectively. It is clear that the radii estimates of \citet{konopacky10} are overestimated, based on an orthogonally aligned system. Assuming such an alignment, a period of 1.86 $\pm$ 0.02 hours is inconsistent with that of LP 349-25A, which requires a much smaller radius of $\sim$0.96 $R_{J}$. However, a radius of $\sim$1.45 $R_{J}$ is derived here for LP 349-25B, which is in loose agreement with the estimates of \citet{dupuy10}, by taking errors in the period and \textit{v} sin \textit{i} into account. We therefore have a case to tentatively assign the period of a 1.86 $\pm$ 0.02 hours to LP 349-25B, as well as possible spin-orbit alignment for this component of the system.}\label{fig11}
\end{center}
\end{figure}

\begin{figure}[!t]
\begin{center}
\includegraphics[scale=0.58]{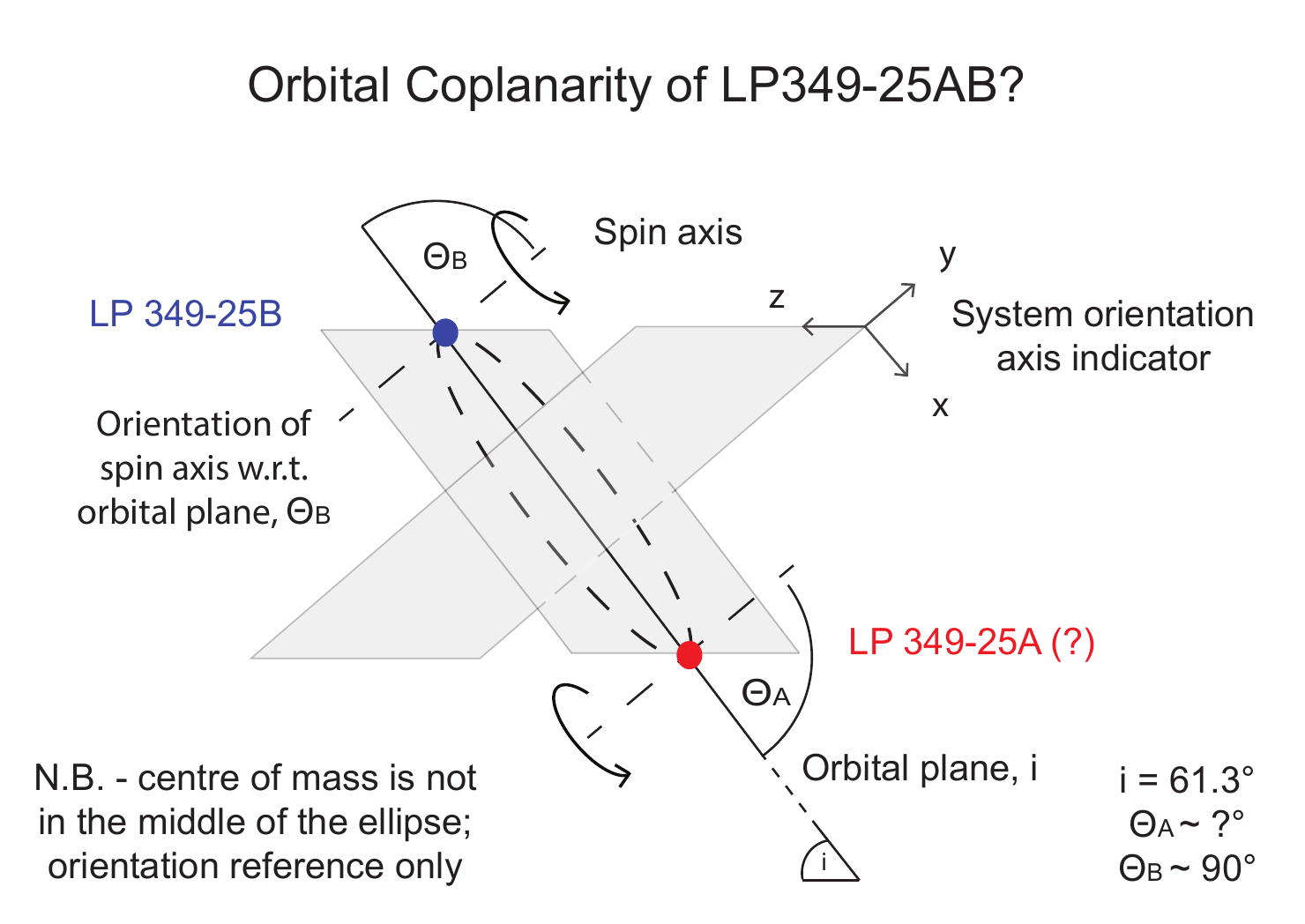}
\caption{A sketch of the configuration of LP 349-25AB, which loosely illustrates the possible system orientation. Based on a radius estimate for LP 349-25B of $\sim$1.37 $R_{J}$ \citep{dupuy10}, in addition to the \textit{v} sin \textit{i} of 83 $\pm$ 3 km s$^{-1}$ \citep{konopacky12}, and the period of 1.86 $\pm$ 0.2 hours in this work, there is tentative evidence that the orientation of the equatorial axis of LP 349-25B, $\Theta_{B}$, is perpendicularly aligned with the binary orbital plane.}\label{fig12}
\end{center}
\end{figure}

\subsubsection{Radius \& inferred spin-orbit alignment}

\par \citet{dupuy10} obtained dynamical mass measurements of a sample of late-M dwarfs, including LP 349-25AB. Their modeling subsequently yield radii estimates of $\sim$1.30 - 1.44 $R_{J}$ for LP 349-25A and $\sim$1.24 - 1.37 $R_{J}$ for LP 349-25B. However, \citet{konopacky10} find much larger radii estimates of 1.7$^{+0.08}_{-0.09}$ $R_{J}$ (A) and 1.68$^{+0.09}_{-0.08}$ $R_{J}$ (B). These studies based their radii on evolutionary model-derived parameters \citep{burrows97,chabrier00,allard01}. Under the assumption of a perfectly coplanar spin-orbit alignment, by adopting the individual rotational velocity measurements of \citet{konopacky12} and by assigning the detected period in this work of 1.86 $\pm$ 0.02 hours to each component, we derive radii of $\sim$0.96 $R_{J}$ for LP 349-25A, and $\sim$1.45 $R_{J}$ for LP 349-25B. We show these in Figure~\ref{fig11} by the dash-dotted horizontal lines, where we have plotted the system's equatorial velocity vs. inclination angle (refer to caption). 

\par Considering the radii estimates of \citet{konopacky10}, as well as an orbital inclination angle of 61.3 $\pm$ 1.5 degrees from their work, we derive a maximum period of rotation of $\sim$3.77 hours and $\sim$2.47 hours for LP 349-25A and LP 349-25B, respectively. Indeed, these radii estimates appear to be very large when considering the evolutionary models of \citet{chabrier00} for a given range of ages, $L_{bol}$, and total system mass presented in their work, in addition to a lack of detected lithium in the binary spectra \citep{reiners09}. Therefore, it is difficult to infer which component matches our detected period.

\par As previously noted, the \citet{dupuy10} binary radii instead infer maximum periods of $\sim$2.65 hours and $\sim$1.67 hours respectively. This discrepancy could be due to the fact that \citet{konopacky10} use only broadband photometry, and furthermore use the effective temperature as one of the inputs for model-predicted mass, whereas \citet{dupuy10} obtain their temperature estimates via NIR fitting, which is $\sim$650 K higher. Notably, determining an accurate estimate of the radius of young, magnetically active stars can be very difficult based on the effect of a reduction in convective efficiency of such objects ($<$0.35 $M_{\odot}$), see \citet{chabrier07}. Since the adiabatic properties of a star increase with mass, such an environment reduces convection in the outer areas. The end result is a reduction in stellar luminosity and core temperature, causing the radius to expand.

\par  Nevertheless, a radius estimate of $\sim$1.45 $R_{J}$ for LP 349-25B (derived above) is in lose agreement with the estimates of \citet{dupuy10}, and therefore we highlight a tentative spin-orbit alignment for the secondary star - as shown by the sketch in Figure~\ref{fig12}. Establishing the period of rotation of the other binary component will enable a more effective constraint of the orbital properties. Finally, as in the case of 2M J0746AB, the inclinations of the spin axes with respect to our line of sight may be equal, but this does not always imply that the orbital planes are perpendicularly aligned. Even if edge-on systems are orthogonal to the sky, they could be coincidentally equal. We refer the reader to \citet{harding13} for a discussion of the various formation mechanisms, and the implications for formation theory in the VLM binary regime.

\section{SUMMARY AND CONCLUSIONS}\label{conclusions}

\par We have reported on optical photometric observations of six ultracool dwarfs spanning the $\sim$M8 - L3.5 spectral range. Our work has confirmed periodic optical variability for five out of six of the radio active dwarf sample, where periodicity for two of these was discovered for the first time. We report a tentative detection of periodic variability for another dwarf - sampling the rotational phase of this object will establish whether periodic variability is also present. Based on previous surveys that have yielded a low fraction of periodic variability for late-M and L dwarfs, our results indicate a likely correlation between the optical and radio periodic variability. This correlation implies that the optical and radio periodic emissions may be related by some kind of magnetic phenomena; however at this point it is not clear whether such a possible connection is causal in nature.  

\par For one of our targets in particular, the pulsing M9 ultracool dwarf TVLM 513, we find periodic variability that is extremely stable in phase over baselines of $\sim$5 years. We achieved an accurate enough rotation period of 1.95958 $\pm$ 0.00005 hours that allowed us to phase connect the $\sim$5 year baseline. The high level of phase stability indicates that the stellar feature responsible for the periodic variability is not moving over these time scales. We do however observe large changes in the peak to peak amplitude variability, pointing toward changes in the size or intensity of the source regions, responsible for the periodicity. 

\par Similarly, for the M tight binary dwarf LP 349-25, the peak to peak amplitude variations change significantly during observations. The phase also changes on these time scales, where we observe it to move in and out of phase during single nights. A number of scenarios are considered for this behavior - e.g. the presence of two periodically varying sources. These changes in morphology could also be a consequence of a high-dynamic environment in these regions, where features are changing in size, temperature and shape, and/or are moving with respect to the stellar photosphere.

\par Finally, we assess the spin-orbit alignment of LP 349-25, based on the discovery of the rotation period for one component in this work. By adopting the radii estimates of \citet{dupuy10}, and assigning the period of rotation of 1.86 hours discovered here for LP 349-25B, we find evidence for a tentative alignment of the spin-orbital axes. Such an alignment has been observed for another VLM binary dwarf - 2M J0746AB \citep{harding13}. Establishing the second period of rotation for the system would further constrain its orbital properties, and further provide insight into the possible formation mechanisms responsible for such alignments in the VLM binary regime.

\acknowledgments

\par This work was largely carried out under the National University of Ireland Traveling Studentship in the Sciences (Physics). LKH gratefully acknowledges the support of the Science Foundation Ireland (Grant Number 07/RFP/PHYF553). We thank the VATT team for their help and guidance, especially Dave Harvey, Michael Franz and Ken Duffek. LKH would also like to personally thank Dr. Mark Lang for his constant assistance, in addition to the staff of the NRAO who provided significant support when LKH was conducting some of his work during his appointment as an NRAO graduate intern. We extend our thanks to Dr. Stuart Littlefair for his many helpful comments on this work. Finally we would like to thank the editor, Dr. Steven Kawaler, in addition to the referee for their careful reading of our work, and for their helpful input on how to improve this manuscript.

\clearpage




\end{document}